\documentclass[twocolumn,preprint,trackchanges]{aastex63}

\usepackage{tabularx}
\usepackage{amsmath,amssymb}
\usepackage{color}
\usepackage{xcolor}
\usepackage{multirow}
\usepackage{natbib}
\usepackage{multirow} 
\usepackage{color, colortbl}
\usepackage{array,ragged2e,longtable}
\usepackage{CJK}

\hypersetup{linkcolor=cyan,citecolor=magenta,filecolor=yellow,urlcolor=blue}

\hypersetup{linkcolor=magenta, citecolor=cyan, filecolor=yellow, urlcolor=blue}

\received{ddmmyyyy}
\revised{\today}
\accepted{ddmmyyyy}
\published{ddmmyyyy}

\submitjournal{ApJ}

\begin{document}
\begin{CJK*}{UTF8}{gbsn}

\title{Probing electromagnetic-gravitational wave emission coincidence in type I binary-driven hypernova family of long GRBs at very-high redshift}

\author[0000-0001-7749-4078]{C.~L.~Bianco}
\affiliation{ICRANet, Piazza della Repubblica 10, I-65122 Pescara, Italy}
\affiliation{ICRA, Dip. di Fisica, Sapienza Universit\`a  di Roma, Piazzale Aldo Moro 5, I-00185 Roma, Italy}
\affiliation{Universit\'e de Nice Sophia-Antipolis, Grand Ch\^ateau Parc Valrose, Nice, CEDEX 2, France}
\affiliation{INAF, Istituto di Astrofisica e Planetologia Spaziali, Via Fosso del Cavaliere 100, 00133 Rome, Italy}

\author[0000-0001-9277-3366]{M.~T.~Mirtorabi}
\affiliation{ICRANet, Piazza della Repubblica 10, I-65122 Pescara, Italy}
\affiliation{Department of Fundamental Physics, Faculty of Physics, Alzahra University, Tehran, Iran}

\author[0000-0002-2516-5894]{R.~Moradi}
\affiliation{ICRANet, Piazza della Repubblica 10, I-65122 Pescara, Italy}
\affiliation{ICRA, Dip. di Fisica, Sapienza Universit\`a  di Roma, Piazzale Aldo Moro 5, I-00185 Roma, Italy}
\affiliation{INAF -- Osservatorio Astronomico d'Abruzzo, Via M. Maggini snc, I-64100, Teramo, Italy}
\affiliation{Key Laboratory of Particle Astrophysics, Institute of High Energy Physics, Chinese Academy of Sciences, Beijing 100049, People's Republic of China}

\author{F.~Rastegarnia}
\affiliation{ICRANet, Piazza della Repubblica 10, I-65122 Pescara, Italy}
\affiliation{Department of Fundamental Physics, Faculty of Physics, Alzahra University, Tehran, Iran}

\author[0000-0003-4904-0014]{J.~A.~Rueda}
\affiliation{ICRANet, Piazza della Repubblica 10, I-65122 Pescara, Italy}
\affiliation{ICRA, Dip. di Fisica, Sapienza Universit\`a  di Roma, Piazzale Aldo Moro 5, I-00185 Roma, Italy}
\affiliation{ICRANet-Ferrara, Dip. di Fisica e Scienze della Terra, Universit\`a degli Studi di Ferrara, Via Saragat 1, I--44122 Ferrara, Italy}
\affiliation{Dip. di Fisica e Scienze della Terra, Universit\`a degli Studi di Ferrara, Via Saragat 1, I--44122 Ferrara, Italy}
\affiliation{INAF, Istituto di Astrofisica e Planetologia Spaziali, Via Fosso del Cavaliere 100, 00133 Rome, Italy}

\author[0000-0003-0829-8318]{R.~Ruffini}
\affiliation{ICRANet, Piazza della Repubblica 10, I-65122 Pescara, Italy}
\affiliation{ICRA, Dip. di Fisica, Sapienza Universit\`a  di Roma, Piazzale Aldo Moro 5, I-00185 Roma, Italy}
\affiliation{INAF, Viale del Parco Mellini 84, 00136 Rome, Italy}

\author[0000-0001-7959-3387]{Y. ~Wang (王瑜)}
\affiliation{ICRA, Dip. di Fisica, Sapienza Universit\`a  di Roma, Piazzale Aldo Moro 5, I-00185 Roma, Italy}
\affiliation{ICRANet, Piazza della Repubblica 10, I-65122 Pescara, Italy}
\affiliation{INAF -- Osservatorio Astronomico d'Abruzzo, Via M. Maggini snc, I-64100, Teramo, Italy}

\author[0000-0003-3142-5020]{M.~Della~Valle}
\affiliation{INAF - Osservatorio Astronomico di Capodimonte, Salita Moiariello 16, I-80131, Napoli, Italy}
\affiliation{ICRANet, Piazza della Repubblica 10, I-65122 Pescara, Italy}

\author[0000-0002-1343-3089]{Liang~Li (李亮)}
\affiliation{ICRANet, Piazza della Repubblica 10, I-65122 Pescara, Italy}
\affiliation{ICRA, Dip. di Fisica, Sapienza Universit\`a  di Roma, Piazzale Aldo Moro 5, I-00185 Roma, Italy}
\affiliation{INAF -- Osservatorio Astronomico d'Abruzzo, Via M. Maggini snc, I-64100, Teramo, Italy}

\author{S.~R.~Zhang (张书瑞)}
\affiliation{ICRANet, Piazza della Repubblica 10, I-65122 Pescara, Italy}
\affiliation{ICRANet-Ferrara, Dip. di Fisica e Scienze della Terra, Universit\`a degli Studi di Ferrara, Via Saragat 1, I--44122 Ferrara, Italy}
\affiliation{School of Astronomy and Space Science, University of Science and Technology of China, Hefei 230026, China }
\affiliation{CAS Key Laboratory for Research in Galaxies and Cosmology, Department of Astronomy, University of Science and Technology of China, Hefei 230026, China}

\email{ruffini@icra.it}

\begin{abstract}
The repointing time of the XRT instrument on the Neil Gehrels Swift Observatory satellite has posed challenges in observing and studying the early X-ray emissions within $\approx40$~s after a gamma-ray burst (GRB) trigger. To address this issue, we adopt a novel approach that capitalizes on the cosmological time dilation in GRBs with redshifts ranging from $3$ to $9$. Applying this strategy to Swift/XRT data, we investigate the earliest X-ray emissions of $368$~GRBs from the Swift catalog, including short and long GRBs. We compare the time delay between the GRB trigger and the initial Swift/XRT observation, measured in the GRB observer frame (OTD) and the cosmological rest-frame (RTD). This technique is here used in the analysis of GRB~090423 at $z=8.233$ (RTD~$\sim8.2$~s), GRB~090429B at $z\approx9.4$ (RTD~$\sim10.1$~s), and GRB~220101A at $z=4.61$ (RTD~$\sim14.4$~s). The cosmological time dilation enables us to observe the very early X-ray afterglow emission in these three GRBs. We thus validate the observation of the collapse of the carbon-oxygen (CO) core and the coeval newborn neutron star ($\nu$NS) formation triggering the GRB event in the binary-driven hypernova (BdHN) scenario. We also evidence the $\nu$NS spin-up due to supernova ejecta fallback and its subsequent slowing down due to the X-optical-radio synchrotron afterglow emission. A brief gravitational wave signal may separate the two stages due to a fast-spinning $\nu$NS triaxial-to-axisymmetric transition. We also analyze the long GRB redshift distribution for the different BdHN types and infer that BdHNe~II and III may originate the NS binary progenitors of short GRBs.
\end{abstract}


\section{Introduction} \label{sec:intro}

Important astronomical breakthroughs are often marked by the possibility of studying events occurring in the nearby Universe. There are several prominent examples, e.g., supernova SN~1987A: its proximity has allowed the first detection of neutrinos \citep{1987PhRvL..58.1490H,1988PhLB..205..209A,1987PhRvL..58.1494B} and the observation of the shock-breakout \citep{1989ARA&A..27..629A}. Another example is gamma-ray burst GRB~980425 and SN~1998bw, the prototype of GRB-SN connection \citep{1998Natur.395..670G,2001ApJ...555..900P}, which occurred at about $40$~Mpc. It is still the closest case of GRB-SN connection observed so far. The problem of the GRB-SN connection has been addressed in \citet{2023ApJ...955...93A}, where the binary-driven hypernova (BdHN) model has been illustrated (see also Appendix~\ref{app:A} in this article). Again, the most important information has been gained from nearby sources (GRB~190114C with $z=0.0425$ and $E_{\rm iso} = 2.5 \times 10^{53}$~erg, GRB~171205A with $z=0.0368$ and $E_{\rm iso} = 5.7 \times 10^{49}$~erg, GRB~190829A with $z=0.0785$ and $E_{\rm iso} = 2\times 10^{50}$~erg). Unlike the cases briefly illustrated above, in this work, we show how the observation of GRBs at very high redshift, by exploiting the cosmological time dilatation factor $(1+z)$ as a novel observational tool, can allow us to enter the \textit{terra incognita} of the very early GRB X-ray emission. This emission is currently inaccessible to the Swift/XRT detector in nearby events, which paradoxically would be more suitable to be studied. However, the significant instrumental delay of repointing the Swift/XRT detector following the GRB trigger, expressed in the observer's rest frame, prevents their early X-ray emission observations. Cosmology, here used as an observational tool, lets us detect the GRB early X-ray afterglow emission at very high redshift. This is one of the main results presented in this article.

For this task, we use the complete sample of $368$ GRBs of the Swift GRB database (see \url{https://swift.gsfc.nasa.gov/archive/grb_table/}) with a measured redshift (see Sec.~\ref{sec:sample}). The sample includes short and long GRBs. We define the observed time delay (OTD) as the time after the GRB trigger needed by Swift/XRT to repoint the source measured in the observer frame\footnote{Namely, the column ``XRT Time to First Observation [sec]'' in the Swift GRB catalog, see \url{https://swift.gsfc.nasa.gov/archive/grb_table/}.} (for details see Sec.~\ref{sec:delay} and, e.g., E. Troja, ``The Neil Gehrels Swift Observatory Technical Handbook Version 17.0'', \url{https://swift.gsfc.nasa.gov/proposals/tech_appd/swiftta_v17.pdf}, as well as \citealp{2004ApJ...611.1005G}). The minimum OTD in our sample is $43.88$~s for GRB~140206A at redshift $z=2.73$ (marked by a horizontal green line in Fig.~\ref{fig:delay1}). It becomes clear that Swift/XRT is generally technically unable to observe the X-ray emission in the first tens of seconds after the GRB trigger. Hence, the X-ray emission within $\approx 40$~s of the GRB trigger remains unobservable by the only available instrument, Swift/XRT. This time interval represents an uncharted new X-ray territory. This large OTD can be circumvented by considering the cosmological corrections presented in this article, using the cosmological rest-frame time delay (RTD, see Sec.~\ref{sec:delay}).

In Sec.~\ref{sec:protot}, special attention is dedicated to the three prototypes of BdHNe~I: GRB~220101A at $z=4.61$ \citep{2022GCN.31353....1F,2022GCN.31357....1P,2022GCN.31359....1F}; GRB~090423 at $z=8.2$ \citep{2009Natur.461.1258S,2009Natur.461.1254T,2014A&A...569A..39R}; and GRB~090429B at a photometric redshift $z\sim 9.4$ \citep{2011ApJ...736....7C}, duly accounting for the $k$-correction and the $0.3$--$10$~keV luminosity light curves. Their excellent data creates the condition to analyze long GRBs' very early X-ray afterglow emission, which is a stringent test for all theoretical models. This is particularly relevant in the case of the BdHN model since it allows the analysis of the new physics of the SN-rise. This provides a unique opportunity to confirm the observation of Episode~I, which corresponds to the carbon-oxygen (CO) core collapse and to the coeval newborn neutron star ($\nu$NS) formation, both of which trigger the onset of the BdHN event. We then evidence the observation of Episode~II. This allows us to identify the physical processes occurring in the $\nu$NS-rise as announced in \citet{2023ApJ...955...93A, 2022PhRvD.106h3004R}, see Sec.~\ref{sec:protot}.

Sec.~\ref{sec:6} discusses the implications of the above findings for the distribution of BHs across the Universe.

In Sec.\ref{sec:subsample}, we finally analyze the redshift distributions of the long GRBs belonging to different BdHN families. We show that their distribution supports the BdHN theoretical prediction, presented in  \citet{2015PhRvL.115w1102F,2016ApJ...832..136R,2018ApJ...859...30R}, that BdHNe~II and III may form the neutron star (NS) binary progenitors of short GRBs. 

Finally, we summarize our conclusions in Sec.~\ref{sec:7}.

\section{Our sample of 368 GRBs and their redshift distribution}\label{sec:sample}

For the analysis of the cosmological time dilation, we build our GRB sample including all GRBs (long and short) that respect these three criteria:
\begin{enumerate}
\item The GRB is present in the Swift GRB database (see \url{https://swift.gsfc.nasa.gov/archive/grb_table/}).
\item The GRB has a measured redshift reported in the Swift GRB database.
\item The GRB has XRT observations with a measured delay between the GRB trigger time determined by Swift/BAT and the moment of the first Swift/XRT observation.
\end{enumerate}
This sample comprises $368$~GRBs from 2005 to the end of 2023 (see Table~\ref{tab:list-XRT}).

\startlongtable
\begin{deluxetable}{ccccc}
\tablewidth{0pt}
\tabletypesize{\scriptsize}
\tablecaption{List of GRBs observed by Swift/XRT and their observed time delay (OTD) and cosmological rest-frame time delay (RTD) in seconds. The delay time is between the initial burst detection and the start time of the first XRT observation. The XRT start time data is sourced from \url{https://swift.gsfc.nasa.gov/archive/grb_table/}. The bold GRB names in this table indicate GRBs with an RTD of less than $43.9$~s, namely shorter than the minimum OTD.\label{tab:list-XRT}}
\tablehead{
\colhead{\#}
&\colhead{GRB}
&\colhead{redshift}
&\colhead{OTD (s)}
&\colhead{RTD (s)}
}
\startdata
\hline
1 & 231215A & 2.305 & 2725.7 & 824.7 \\
2 & \textbf{231210B} & 3.13 & 123.4 & 29.9 \\
3 & 231118A & 0.8304 & 413.8 & 226.1 \\
4 & 231117A & 0.257 & 101.8 & 81 \\
5 & \textbf{231111A} & 1.28 & 99.6 & 43.7 \\
6 & \textbf{230818A} & 2.42 & 141.1 & 41.3 \\
7 & 230506C & 3.7 & 4600 & 978.7 \\
8 & \textbf{230414B} & 3.568 & 128.6 & 28.1 \\
9 & 230328B & 0.09 & 110.3 & 101.2 \\
10 & \textbf{230325A} & 1.664 & 94.1 & 35.3 \\
11 & \textbf{230116D} & 3.81 & 130.9 & 27.2 \\
12 & \textbf{221226B} & 2.694 & 102.9 & 27.9 \\
13 & \textbf{221110A} & 4.06 & 53.1 & 10.5 \\
14 & 221009A & 0.151 & 91.6 & 79.6 \\
15 & 220611A & 2.3608 & 149.2 & 44.4 \\
16 & \textbf{220521A} & 5.6 & 96.3 & 14.6 \\
17 & \textbf{220117A} & 4.961 & 151.9 & 25.5 \\
18 & 220107A & 1.246 & 26700 & 11887.8 \\
19 & \textbf{220101A} & 4.61 & 80.8 & 14.4 \\
20 & \textbf{211207A} & 2.272 & 80.5 & 24.6 \\
21 & 211024B & 1.1137 & 105 & 49.7 \\
22 & 211023B & 0.862 & 95.3 & 51.2 \\
23 & \textbf{210905A} & 6.318 & 91.7 & 12.5 \\
24 & \textbf{210822A} & 1.736 & 74.6 & 27.3 \\
25 & 210731A & 1.2525 & 200.9 & 89.2 \\
26 & \textbf{210722A} & 1.145 & 84.8 & 39.6 \\
27 & 210702A & 1.1757 & 95.5 & 43.9 \\
28 & 210619B & 1.937 & 328.1 & 111.7 \\
29 & \textbf{210610B} & 1.13 & 83.9 & 39.4 \\
30 & \textbf{210610A} & 3.54 & 90 & 19.8 \\
31 & \textbf{210517A} & 2.486 & 67.1 & 19.2 \\
32 & 210504A & 2.077 & 218 & 70.8 \\
33 & 210420B & 1.4 & 141 & 58.8 \\
34 & \textbf{210411C} & 2.826 & 63.1 & 16.5 \\
35 & 210321A & 1.487 & 2730.6 & 1098 \\
36 & \textbf{210222B} & 2.198 & 95.9 & 30 \\
37 & 210210A & 0.715 & 82.1 & 47.9 \\
38 & \textbf{201221D} & 1.046 & 87.4 & 42.7 \\
39 & \textbf{201221A} & 5.7 & 136.5 & 20.4 \\
40 & 201216C & 1.1 & 2966.8 & 1412.8 \\
41 & \textbf{201104B} & 1.954 & 102 & 34.5 \\
42 & \textbf{201024A} & 0.999 & 74.9 & 37.5 \\
43 & 201021C & 1.07 & 102 & 49.3 \\
44 & \textbf{201020A} & 2.903 & 141.5 & 36.3 \\
45 & 201015A & 0.426 & 3214.1 & 2253.9 \\
46 & \textbf{201014A} & 4.56 & 156 & 28.1 \\
47 & 200829A & 1.25 & 128.7 & 57.2 \\
48 & 200522A & 0.4 & 83.4 & 59.6 \\
49 & 200205B & 1.465 & 342.7 & 139 \\
50 & \textbf{191221B} & 1.169 & 86.3 & 39.8 \\
51 & 191019A & 0.248 & 3210.9 & 2572.9 \\
52 & \textbf{191011A} & 1.722 & 74.8 & 27.5 \\
53 & \textbf{191004B} & 3.503 & 57.7 & 12.8 \\
54 & 190829A & 0.0785 & 97.3 & 90.2 \\
55 & \textbf{190719C} & 2.469 & 60.9 & 17.6 \\
56 & \textbf{190627A} & 1.942 & 109.8 & 37.3 \\
57 & 190324A & 1.1715 & 3297.9 & 1518.7 \\
58 & 190114C & 0.42 & 64 & 45 \\
59 & 190114A & 3.3765 & 246.6 & 56.4 \\
60 & \textbf{190106A} & 1.86 & 81.8 & 28.6 \\
61 & 181213A & 2.4 & 2342.8 & 689.1 \\
62 & \textbf{181110A} & 1.505 & 64 & 25.5 \\
63 & \textbf{181020A} & 2.938 & 55.6 & 14.1 \\
64 & \textbf{181010A} & 1.39 & 93.1 & 38.9 \\
65 & 180728A & 0.117 & 1730.8 & 1549.5 \\
66 & 180720B & 0.654 & 86.5 & 52.3 \\
67 & \textbf{180624A} & 2.855 & 112.2 & 29.1 \\
68 & \textbf{180620B} & 1.1175 & 83.2 & 39.3 \\
69 & 180510B & 1.305 & 2950.1 & 1279.9 \\
70 & \textbf{180404A} & 1 & 86.5 & 43.2 \\
71 & \textbf{180329B} & 1.998 & 103.6 & 34.6 \\
72 & \textbf{180325A} & 2.25 & 73.4 & 22.6 \\
73 & 180314A & 1.445 & 159.3 & 65.1 \\
74 & \textbf{180115A} & 2.487 & 131.1 & 37.6 \\
75 & 171222A & 2.409 & 169.7 & 49.8 \\
76 & 171205A & 0.0368 & 144.7 & 139.6 \\
77 & 171020A & 1.87 & 144.5 & 50.3 \\
78 & 170903A & 0.886 & 3117.5 & 1653 \\
79 & 170714A & 0.793 & 392.7 & 219 \\
80 & \textbf{170705A} & 2.01 & 72.3 & 24 \\
81 & 170607A & 0.557 & 73.4 & 47.1 \\
82 & 170604A & 1.329 & 124.7 & 53.6 \\
83 & \textbf{170531B} & 2.366 & 140.9 & 41.9 \\
84 & 170519A & 0.818 & 80.4 & 44.2 \\
85 & \textbf{170405A} & 3.51 & 120.6 & 26.7 \\
86 & \textbf{170202A} & 3.645 & 72.5 & 15.6 \\
87 & \textbf{170113A} & 1.968 & 58.7 & 19.8 \\
88 & 161219B & 0.1475 & 108.3 & 94.3 \\
89 & 161129A & 0.645 & 82 & 49.8 \\
90 & \textbf{161117A} & 1.549 & 60.8 & 23.9 \\
91 & \textbf{161108A} & 1.159 & 80.3 & 37.2 \\
92 & \textbf{161017A} & 2.0127 & 57.7 & 19.1 \\
93 & \textbf{161014A} & 2.823 & 121.8 & 31.9 \\
94 & 160804A & 0.736 & 147 & 84.7 \\
95 & 160703A & 1.5 & 107800 & 43120 \\
96 & 160624A & 0.483 & 73.7 & 49.7 \\
97 & 160425A & 0.555 & 203.4 & 130.8 \\
98 & \textbf{160410A} & 1.717 & 82.9 & 30.5 \\
99 & \textbf{160327A} & 4.99 & 60.5 & 10.1 \\
100 & 160314A & 0.726 & 91 & 52.7 \\
101 & 160227A & 2.38 & 151.9 & 44.9 \\
102 & \textbf{160203A} & 3.52 & 137.3 & 30.4 \\
103 & \textbf{160131A} & 0.97 & 69.7 & 35.4 \\
104 & \textbf{160121A} & 1.96 & 91.1 & 30.8 \\
105 & \textbf{160117B} & 0.87 & 55 & 29.4 \\
106 & 151215A & 2.59 & 169.1 & 47.1 \\
107 & 151112A & 4.1 & 3141 & 615.9 \\
108 & \textbf{151111A} & 3.5 & 72.5 & 16.1 \\
109 & 151031A & 1.167 & 433.6 & 200.1 \\
110 & \textbf{151027B} & 4.063 & 203.4 & 40.2 \\
111 & 151027A & 0.38 & 87 & 63 \\
112 & \textbf{151021A} & 2.33 & 90.7 & 27.2 \\
113 & \textbf{150915A} & 1.968 & 128.7 & 43.4 \\
114 & 150910A & 1.359 & 145.3 & 61.6 \\
115 & 150821A & 0.755 & 243.3 & 138.6 \\
116 & 150818A & 0.282 & 84.5 & 65.9 \\
117 & 150727A & 0.313 & 77.2 & 58.8 \\
118 & \textbf{150424A} & 3 & 87.9 & 22 \\
119 & \textbf{150423A} & 1.394 & 70.1 & 29.3 \\
120 & 150413A & 3.2 & 303300 & 72214.3 \\
121 & \textbf{150403A} & 2.06 & 74.7 & 24.4 \\
122 & 150323A & 0.593 & 146.6 & 92 \\
123 & \textbf{150314A} & 1.758 & 85.1 & 30.9 \\
124 & \textbf{150301B} & 1.5169 & 82.4 & 32.8 \\
125 & 150206A & 2.087 & 474.5 & 153.7 \\
126 & 150120A & 0.46 & 76.2 & 52.2 \\
127 & 150101B & 0.093 & 139200 & 127355.9 \\
128 & 141225A & 0.915 & 423.5 & 221.2 \\
129 & \textbf{141221A} & 1.452 & 79.5 & 32.4 \\
130 & \textbf{141220A} & 1.3195 & 99.2 & 42.8 \\
131 & \textbf{141212A} & 0.596 & 69.1 & 43.3 \\
132 & 141121A & 1.47 & 362.4 & 146.7 \\
133 & \textbf{141109A} & 2.993 & 129.2 & 32.4 \\
134 & \textbf{141026A} & 3.35 & 157 & 36.1 \\
135 & \textbf{141004A} & 0.57 & 59.9 & 38.1 \\
136 & \textbf{140907A} & 1.21 & 83.6 & 37.8 \\
137 & \textbf{140903A} & 0.351 & 59 & 43.7 \\
138 & 140710A & 0.558 & 98.4 & 63.2 \\
139 & \textbf{140703A} & 3.14 & 112.8 & 27.3 \\
140 & \textbf{140629A} & 2.275 & 94.3 & 28.8 \\
141 & 140622A & 0.959 & 93.4 & 47.7 \\
142 & \textbf{140614A} & 4.233 & 123.3 & 23.6 \\
143 & \textbf{140518A} & 4.707 & 69 & 12.1 \\
144 & \textbf{140515A} & 6.32 & 75.8 & 10.4 \\
145 & 140512A & 0.725 & 98.4 & 57 \\
146 & 140506A & 0.889 & 97.9 & 51.8 \\
147 & \textbf{140430A} & 1.6 & 50.8 & 19.5 \\
148 & 140423A & 3.26 & 2943.5 & 691 \\
149 & \textbf{140419A} & 3.956 & 86.5 & 17.5 \\
150 & 140318A & 1.02 & 124.8 & 61.8 \\
151 & 140311A & 4.95 & 9500 & 1596.6 \\
152 & \textbf{140304A} & 5.283 & 75.2 & 12 \\
153 & \textbf{140301A} & 1.416 & 86.1 & 35.6 \\
154 & 140213A & 1.2076 & 3425.3 & 1551.6 \\
155 & \textbf{140206A} & 2.73 & 43.9 & 11.8 \\
156 & 140114A & 3 & 577.5 & 144.4 \\
157 & \textbf{131227A} & 5.3 & 57 & 9 \\
158 & \textbf{131117A} & 4.11 & 66.1 & 12.9 \\
159 & 131105A & 1.686 & 290.9 & 108.3 \\
160 & 131103A & 0.599 & 76.3 & 47.7 \\
161 & \textbf{131030A} & 1.293 & 78.4 & 34.2 \\
162 & \textbf{131004A} & 0.717 & 70 & 40.8 \\
163 & 130925A & 0.347 & 147.4 & 109.4 \\
164 & \textbf{130907A} & 1.238 & 66.6 & 29.8 \\
165 & 130831A & 0.4791 & 125.8 & 85.1 \\
166 & \textbf{130701A} & 1.155 & 85.5 & 39.7 \\
167 & \textbf{130612A} & 2.006 & 87 & 28.9 \\
168 & \textbf{130610A} & 2.092 & 133 & 43 \\
169 & \textbf{130606A} & 5.91 & 72.4 & 10.5 \\
170 & 130604A & 1.06 & 99.3 & 48.2 \\
171 & \textbf{130603B} & 0.356 & 59.1 & 43.5 \\
172 & \textbf{130514A} & 3.6 & 88.8 & 19.3 \\
173 & \textbf{130511A} & 1.3033 & 71.6 & 31.1 \\
174 & \textbf{130505A} & 2.27 & 96.4 & 29.5 \\
175 & \textbf{130427B} & 2.78 & 77.4 & 20.5 \\
176 & 130427A & 0.34 & 140.2 & 104.6 \\
177 & 130420A & 1.297 & 735.3 & 320.1 \\
178 & 130418A & 1.218 & 129.7 & 58.5 \\
179 & \textbf{130408A} & 3.758 & 149.9 & 31.5 \\
180 & \textbf{130131B} & 2.539 & 109.5 & 30.9 \\
181 & \textbf{121229A} & 2.707 & 145.9 & 39.4 \\
182 & 121211A & 1.023 & 89.5 & 44.3 \\
183 & \textbf{121201A} & 3.385 & 115.1 & 26.2 \\
184 & \textbf{121128A} & 2.2 & 77.2 & 24.1 \\
185 & \textbf{121027A} & 1.773 & 67.4 & 24.3 \\
186 & \textbf{121024A} & 2.298 & 93 & 28.2 \\
187 & \textbf{120922A} & 3.1 & 116.4 & 28.4 \\
188 & \textbf{120909A} & 3.93 & 93.4 & 18.9 \\
189 & \textbf{120907A} & 0.97 & 82 & 41.6 \\
190 & 120815A & 2.358 & 2686.8 & 800.1 \\
191 & \textbf{120811C} & 2.671 & 68.7 & 18.7 \\
192 & \textbf{120805A} & 3.1 & 123.1 & 30 \\
193 & \textbf{120802A} & 3.796 & 84.8 & 17.7 \\
194 & \textbf{120729A} & 0.8 & 68.1 & 37.8 \\
195 & 120724A & 1.48 & 109.1 & 44 \\
196 & 120722A & 0.9586 & 153 & 78.1 \\
197 & 120714B & 0.3984 & 120.1 & 85.9 \\
198 & \textbf{120712A} & 4.11 & 90.9 & 17.8 \\
199 & \textbf{120521C} & 6 & 69.1 & 9.9 \\
200 & 120422A & 0.28 & 95.1 & 74.3 \\
201 & \textbf{120404A} & 2.876 & 130 & 33.5 \\
202 & \textbf{120327A} & 2.81 & 75.6 & 19.8 \\
203 & \textbf{120326A} & 1.798 & 59.5 & 21.3 \\
204 & \textbf{120119A} & 1.728 & 53.3 & 19.5 \\
205 & \textbf{120118B} & 2.943 & 112.1 & 28.4 \\
206 & \textbf{111229A} & 1.3805 & 83.3 & 35 \\
207 & 111228A & 0.714 & 145.1 & 84.6 \\
208 & 111225A & 0.297 & 88.1 & 68 \\
209 & 111209A & 0.677 & 418.9 & 249.8 \\
210 & \textbf{110503A} & 1.613 & 93.6 & 35.8 \\
211 & 110422A & 1.77 & 814.5 & 294 \\
212 & \textbf{110213A} & 1.46 & 91.7 & 37.3 \\
213 & 110205A & 1.98 & 155.4 & 52.1 \\
214 & \textbf{110128A} & 2.339 & 140.5 & 42.1 \\
215 & 101225A & 0.623 & 1383 & 852.2 \\
216 & 101219B & 0.5519 & 542.7 & 349.7 \\
217 & 101219A & 0.718 & 221.9 & 129.2 \\
218 & \textbf{100906A} & 1.727 & 80.2 & 29.4 \\
219 & 100902A & 4.5 & 316.2 & 57.5 \\
220 & 100901A & 1.408 & 157 & 65.2 \\
221 & 100816A & 0.8034 & 82.9 & 45.9 \\
222 & \textbf{100814A} & 1.44 & 87.3 & 35.8 \\
223 & \textbf{100728B} & 2.106 & 97.1 & 31.2 \\
224 & \textbf{100728A} & 1.567 & 76.7 & 29.9 \\
225 & \textbf{100724A} & 1.288 & 88.9 & 38.9 \\
226 & 100621A & 0.542 & 76 & 49.3 \\
227 & \textbf{100615A} & 1.398 & 62.4 & 26 \\
228 & \textbf{100513A} & 4.772 & 126.8 & 22 \\
229 & \textbf{100425A} & 1.755 & 78.8 & 28.6 \\
230 & \textbf{100424A} & 2.465 & 119.8 & 34.6 \\
231 & 100418A & 0.6235 & 79.1 & 48.7 \\
232 & 100316D & 0.014 & 137.7 & 135.8 \\
233 & \textbf{100316B} & 1.18 & 64.1 & 29.4 \\
234 & \textbf{100302A} & 4.813 & 125.5 & 21.6 \\
235 & \textbf{100219A} & 4.5 & 178.6 & 32.5 \\
236 & 091208B & 1.063 & 115.1 & 55.8 \\
237 & 091127 & 0.49 & 3214.6 & 2157.5 \\
238 & \textbf{091109A} & 3.063 & 150.7 & 37.1 \\
239 & \textbf{091029} & 2.752 & 79.9 & 21.3 \\
240 & 091024 & 1.092 & 3192 & 1525.8 \\
241 & \textbf{091020} & 1.71 & 81.5 & 30.1 \\
242 & \textbf{091018} & 0.971 & 61.5 & 31.2 \\
243 & 090927 & 1.37 & 2137 & 901.7 \\
244 & \textbf{090926B} & 1.24 & 88.8 & 39.6 \\
245 & 090814A & 1.448 & 159.3 & 65.1 \\
246 & \textbf{090812} & 2.452 & 76.8 & 22.3 \\
247 & \textbf{090809} & 2.737 & 104 & 27.8 \\
248 & 090726 & 2.71 & 3061.7 & 825.3 \\
249 & \textbf{090715B} & 3 & 46.3 & 11.6 \\
250 & 090618 & 0.54 & 120.9 & 78.5 \\
251 & 090529 & 2.625 & 197.1 & 54.4 \\
252 & \textbf{090519} & 3.9 & 114.9 & 23.5 \\
253 & \textbf{090516A} & 4.109 & 170 & 33.3 \\
254 & 090510 & 0.903 & 94.1 & 49.4 \\
255 & \textbf{090429B} & 9.4 & 104.69 & 10.07 \\
256 & \textbf{090426} & 2.609 & 84.6 & 23.4 \\
257 & 090424 & 0.544 & 84.5 & 54.7 \\
258 & \textbf{090423} & 8 & 72.5 & 8.1 \\
259 & \textbf{090418A} & 1.608 & 96.1 & 36.8 \\
260 & \textbf{090407} & 1.4485 & 93 & 38 \\
261 & \textbf{090205} & 4.7 & 87.6 & 15.4 \\
262 & \textbf{090113} & 1.7493 & 70.9 & 25.8 \\
263 & 090102 & 1.547 & 387.2 & 152 \\
264 & \textbf{081222} & 2.7 & 51.8 & 14 \\
265 & \textbf{081221} & 2.26 & 68.4 & 21 \\
266 & \textbf{081203A} & 2.1 & 83.1 & 26.8 \\
267 & 081121 & 2.512 & 2813.2 & 801 \\
268 & \textbf{081118} & 2.58 & 153.3 & 42.8 \\
269 & 081029 & 3.8479 & 2702.9 & 557.5 \\
270 & 081028A & 3.038 & 190.7 & 47.2 \\
271 & \textbf{081008} & 1.9685 & 87.2 & 29.4 \\
272 & 081007 & 0.5295 & 99.4 & 65 \\
273 & 080928 & 1.692 & 169.7 & 63 \\
274 & \textbf{080916A} & 0.689 & 70.2 & 41.6 \\
275 & \textbf{080913} & 6.44 & 99.5 & 13.4 \\
276 & \textbf{080906} & 2 & 71.3 & 23.8 \\
277 & \textbf{080905B} & 2.374 & 103.2 & 30.6 \\
278 & 080905A & 0.1218 & 130.4 & 116.2 \\
279 & \textbf{080810} & 3.35 & 76 & 17.5 \\
280 & \textbf{080804} & 2.2045 & 99 & 30.9 \\
281 & \textbf{080721} & 2.602 & 108 & 30 \\
282 & 080710 & 0.845 & 3131.6 & 1697.3 \\
283 & \textbf{080707} & 1.23 & 68.3 & 30.6 \\
284 & \textbf{080607} & 3.036 & 82.1 & 20.3 \\
285 & \textbf{080605} & 1.6398 & 90.4 & 34.2 \\
286 & 080604 & 1.416 & 119.3 & 49.4 \\
287 & \textbf{080603B} & 2.69 & 61.8 & 16.7 \\
288 & \textbf{080520} & 1.545 & 99.5 & 39.1 \\
289 & \textbf{080516} & 3.2 & 82.9 & 19.7 \\
290 & \textbf{080430} & 0.76 & 48.9 & 27.8 \\
291 & 080413B & 1.1 & 131.3 & 62.5 \\
292 & \textbf{080413A} & 2.433 & 60.7 & 17.7 \\
293 & \textbf{080411} & 1.03 & 70.2 & 34.6 \\
294 & \textbf{080330} & 1.51 & 70.5 & 28.1 \\
295 & 080319C & 1.95 & 223.7 & 75.8 \\
296 & \textbf{080319B} & 0.937 & 60.5 & 31.2 \\
297 & \textbf{080310} & 2.4266 & 89.2 & 26 \\
298 & \textbf{080210} & 2.641 & 157.1 & 43.2 \\
299 & \textbf{080207} & 2.0858 & 124.1 & 40.2 \\
300 & 071227 & 0.383 & 79.1 & 57.2 \\
301 & 071122 & 1.14 & 139.8 & 65.3 \\
302 & 071117 & 1.331 & 2848 & 1221.8 \\
303 & 071112C & 0.823 & 83.6 & 45.9 \\
304 & \textbf{071031} & 2.692 & 102.8 & 27.8 \\
305 & \textbf{071021} & 2.452 & 130.5 & 37.8 \\
306 & \textbf{071020} & 2.145 & 61.2 & 19.5 \\
307 & 071010B & 0.947 & 92631 & 47576.3 \\
308 & \textbf{070802} & 2.45 & 137.9 & 40 \\
309 & 070724A & 0.457 & 66.8 & 45.8 \\
310 & \textbf{070714B} & 0.92 & 61.4 & 32 \\
311 & 070611 & 2.04 & 3287.2 & 1081.3 \\
312 & \textbf{070529} & 2.4996 & 131 & 37.4 \\
313 & 070521 & 0.553 & 76.9 & 49.5 \\
314 & \textbf{070508} & 0.82 & 75.9 & 41.7 \\
315 & \textbf{070506} & 2.31 & 127 & 38.4 \\
316 & 070429B & 0.904 & 256.3 & 134.6 \\
317 & 070419A & 0.97 & 112.9 & 57.3 \\
318 & \textbf{070411} & 2.954 & 96.5 & 24.4 \\
319 & \textbf{070318} & 0.836 & 63.6 & 34.6 \\
320 & 070306 & 1.497 & 153.2 & 61.4 \\
321 & 070208 & 1.165 & 115.5 & 53.3 \\
322 & \textbf{070129} & 2.3384 & 133.7 & 40 \\
323 & \textbf{070110} & 2.352 & 93.4 & 27.9 \\
324 & \textbf{070103} & 2.6208 & 68.6 & 19 \\
325 & \textbf{061222B} & 3.355 & 145.1 & 33.3 \\
326 & \textbf{061222A} & 2.088 & 101 & 32.7 \\
327 & \textbf{061217} & 0.827 & 64 & 35 \\
328 & 061201 & 0.111 & 81.3 & 73.2 \\
329 & \textbf{061121} & 1.314 & 55.4 & 23.9 \\
330 & 061110B & 3.44 & 3042.2 & 685.2 \\
331 & \textbf{061110A} & 0.758 & 69.2 & 39.4 \\
332 & 061021 & 0.3463 & 72.8 & 54.1 \\
333 & \textbf{061007} & 1.261 & 80.5 & 35.6 \\
334 & \textbf{060927} & 5.6 & 64.7 & 9.8 \\
335 & \textbf{060926} & 3.208 & 60 & 14.2 \\
336 & 060912A & 0.937 & 108.9 & 56.2 \\
337 & \textbf{060908} & 1.8836 & 71.7 & 24.9 \\
338 & \textbf{060906} & 3.685 & 148.5 & 31.7 \\
339 & \textbf{060904B} & 0.703 & 68.8 & 40.4 \\
340 & \textbf{060814} & 0.84 & 71.5 & 38.9 \\
341 & 060729 & 0.54 & 124.4 & 80.8 \\
342 & 060719 & 1.532 & 128.8 & 50.9 \\
343 & \textbf{060714} & 2.71 & 99 & 26.7 \\
344 & \textbf{060708} & 2.3 & 62.3 & 18.9 \\
345 & \textbf{060707} & 3.43 & 120.5 & 27.2 \\
346 & 060614 & 0.13 & 91.4 & 80.9 \\
347 & \textbf{060607A} & 3.082 & 65.2 & 16 \\
348 & \textbf{060605} & 3.8 & 92.4 & 19.2 \\
349 & \textbf{060604} & 2.1357 & 108.8 & 34.7 \\
350 & \textbf{060526} & 3.21 & 73.2 & 17.4 \\
351 & \textbf{060522} & 5.11 & 144.4 & 23.6 \\
352 & 060512 & 0.4428 & 101.8 & 70.5 \\
353 & \textbf{060510B} & 4.9 & 118.8 & 20.1 \\
354 & 060502B & 0.287 & 70.3 & 54.6 \\
355 & \textbf{060502A} & 1.51 & 76.3 & 30.4 \\
356 & \textbf{060418} & 1.49 & 78 & 31.3 \\
357 & \textbf{060223A} & 4.41 & 85.9 & 15.9 \\
358 & 060218 & 0.0331 & 153.1 & 148.2 \\
359 & \textbf{060210} & 3.91 & 95 & 19.3 \\
360 & \textbf{060206} & 4.045 & 58.4 & 11.6 \\
361 & \textbf{060124} & 2.3 & 106.1 & 32.2 \\
362 & \textbf{060116} & 5.3 & 153.5 & 24.4 \\
363 & \textbf{060115} & 3.53 & 112.6 & 24.9 \\
364 & \textbf{060108} & 2.03 & 91.4 & 30.2 \\
365 & 051221A & 0.547 & 88 & 56.9 \\
366 & 051117B & 0.481 & 134.8 & 91 \\
367 & 051109B & 0.08 & 86.2 & 79.8 \\
368 & \textbf{051109A} & 2.346 & 119.7 & 35.8 \\
 \hline
\enddata
\end{deluxetable}

\section{The Swift/XRT observed time delay (OTD) and the cosmological rest-frame time delay (RTD)}\label{sec:delay}

\begin{figure*}
\centering
\includegraphics[width=\hsize]{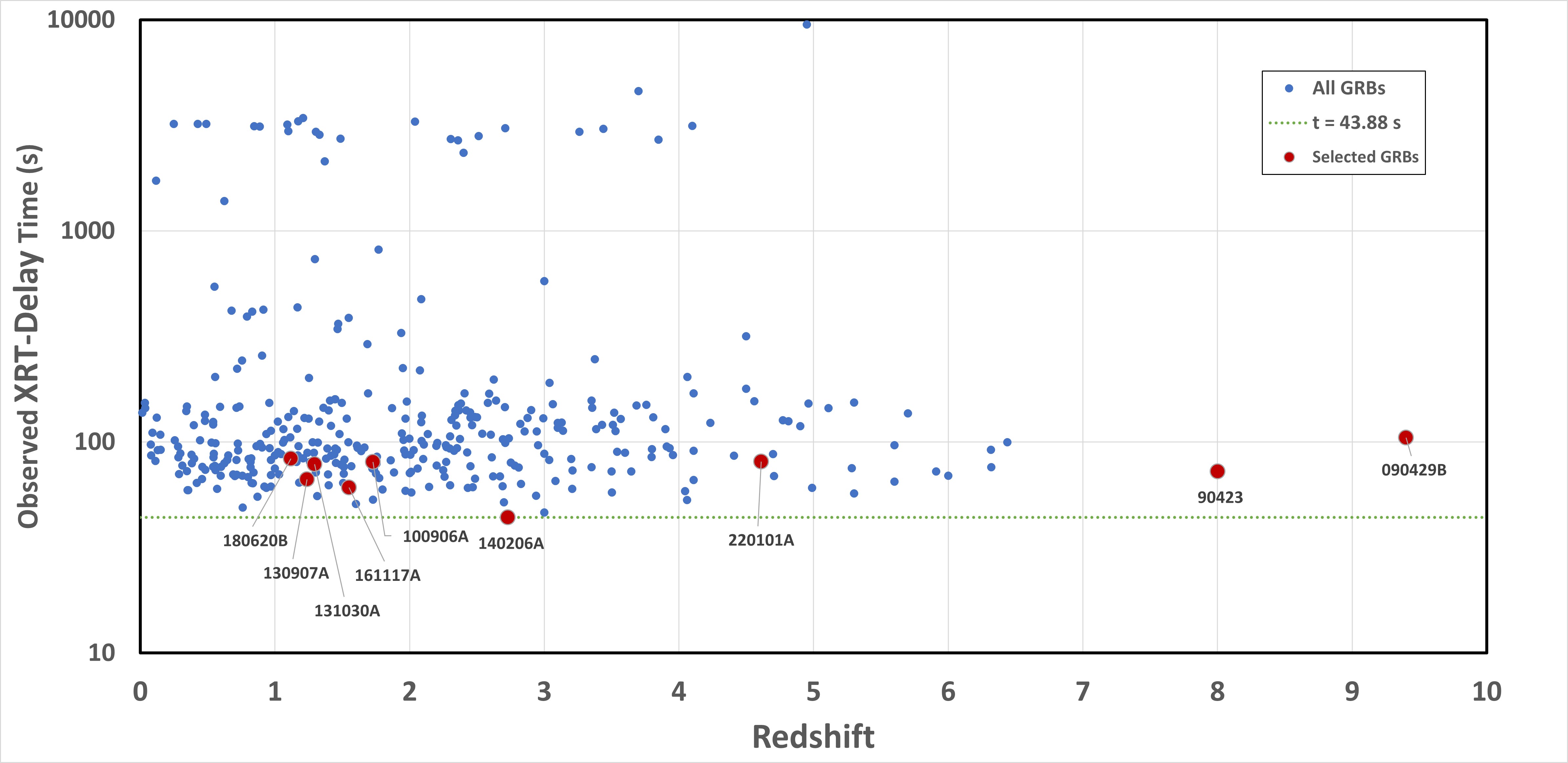}
\includegraphics[width=\hsize]{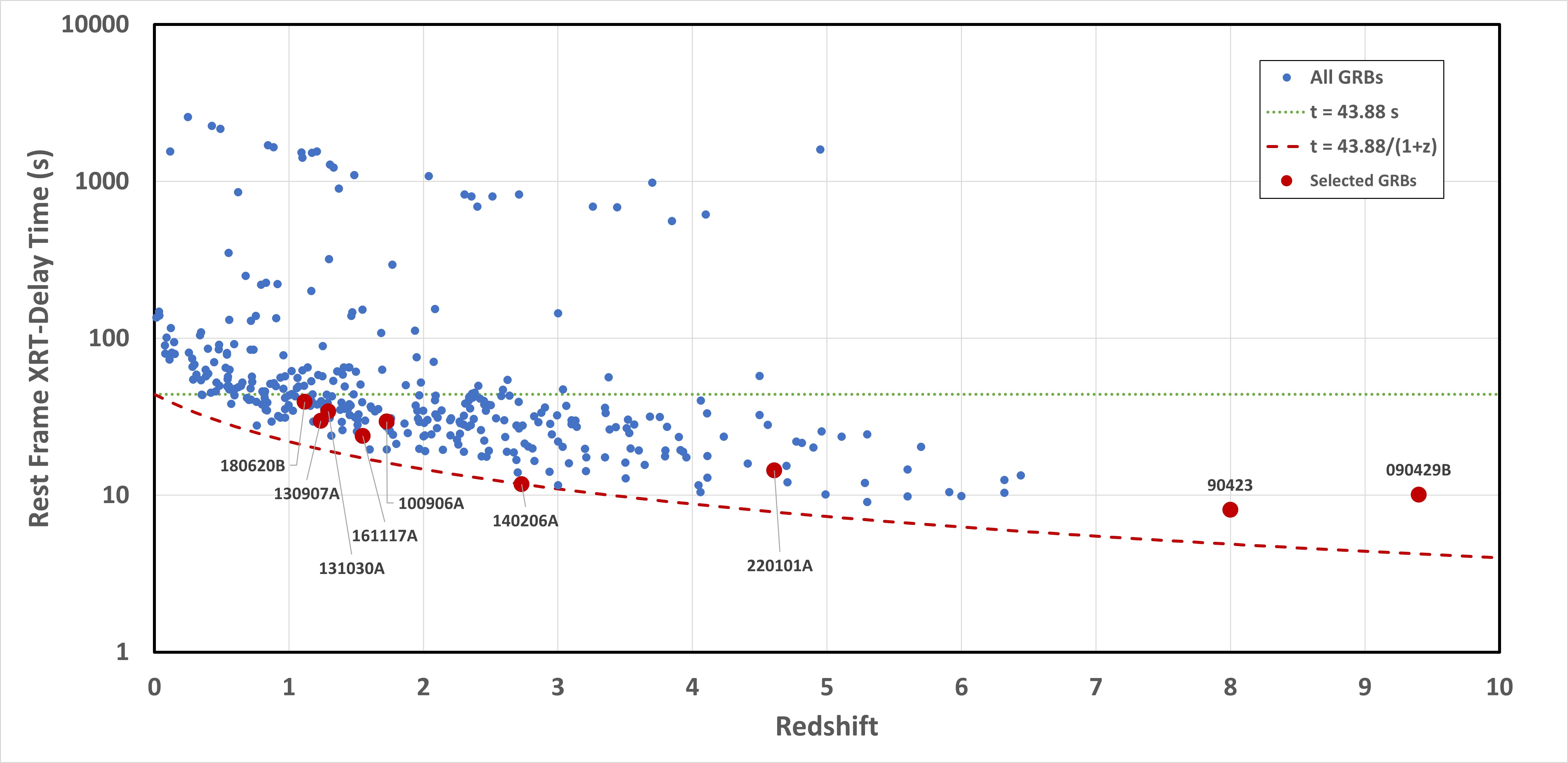}
\caption{The Swift/XRT time delay in the observer's frame (OTD, upper panel) and the cosmological rest-frame (RTD, lower panel). Red circles mark GRB~220101A at $z=4.61$, GRB~090423 at $z=8.2$, GRB~090429B at $z\sim9.4$, GRB~140206A at $z\sim2.73$, GRB~100906A at $z\sim1.727$, GRB~161117 at $z\sim1.549$, GRB~~130907A at $z\sim1.238$, GRB~180620B at $z\sim1.118$ and GRB~131030A at $z\sim1.239$.} \label{fig:delay1}
\end{figure*}

We now turn to the main point of this paper and focus on the Swift/XRT time delays in our sample of $368$~GRBs as a function of their cosmological redshift. We define the observed time delay as the time after the GRB trigger (OTD) needed by Swift/XRT to repoint the source measured in the observer frame\footnote{Namely, the column ``XRT Time to First Observation [sec]'' in the Swift GRB catalog, see \url{https://swift.gsfc.nasa.gov/archive/grb_table/}.} \citep[see][for more information]{2004ApJ...611.1005G}. We plot this quantity in the upper panel of Fig.~\ref{fig:delay1}. The minimum OTD in our sample is $43.88$~s from GRB~140206A at redshift $z=2.73$ (marked by a horizontal green dotted line in the plot). Table~\ref{tab:list-XRT} presents the complete list of the $368$ GRBs in our sample and their OTD in seconds.

It is then clear that Swift/XRT is generally unable to observe the X-ray emission in the first $43$ seconds after the GRB trigger. This is because it takes at least between $10$~s and $20$~s for the Swift satellite to automatically realize that a Swift/BAT trigger condition occurred, to compute the coordinates of the source, to check if a slewing to those coordinates is possible, and to start slewing to put the source in the Swift/XRT field of view; the actual slewing time is between $20$~s and $75$~s (for details see, e.g., E. Troja, ``The Neil Gehrels Swift Observatory Technical Handbook Version 17.0'', \url{https://swift.gsfc.nasa.gov/proposals/tech_appd/swiftta_v17.pdf}, and \citealp{2004ApJ...611.1005G}). Hence, X-ray events occurring within $\approx 40$~s of the GRB trigger remain unobservable by Swift/XRT, making this time interval an X-ray uncharted new territory. Our knowledge during this phase, which corresponds to the prompt emission of GRBs, is confined to fewer than $100$ detections made by BeppoSAX and HETE-2 \citep[see, e.g.,][]{2003ICRC....5.2741T, 2011NCimR..34..585C, 2019RLSFN..30S.171F}.

Interestingly, this large OTD can be circumvented by considering the cosmological corrections presented in this article and turning to the cosmological rest-frame time delay (RTD) in seconds. This procedure has been routinely applied in our approach \citep[see, e.g.,][and references therein]{2021MNRAS.504.5301R}. Due to the cosmological time dilation, a time interval $\Delta t$ measured on Earth corresponds to a time interval $\Delta t/(1+z)$ in the cosmological source rest-frame, where $z$ is its cosmological redshift. In other words, a phenomenon appearing to our instruments on the Earth to last $50$~s may last $10$~s if the source is at $z=4$, like if we were observing the phenomenon in slow motion.

Therefore, the OTD needed by Swift/XRT to start its observations after the GRB trigger may correspond to a much shorter actual RTD for sources with a large redshift $z$, exactly by a factor $(1+z)$. If, e.g., XRT starts to observe a GRB $60$~s after the trigger in the observer frame, it is observing the X-ray signals emitted $60/(1+z)$~s after the trigger in the rest-frame of the source. This corresponds to the possibility of observing $10$~s after the trigger for a GRB with $z=5$: the higher the GRB redshift, the shorter the time Swift/XRT can observe the source after the GRB trigger.

This is clearly shown in the lower panel of Fig.~\ref{fig:delay1}, where we present the time delays of the upper panel converted in the cosmological rest frame of each source; see also Table~\ref{tab:list-XRT} where we compare and contrast OTD and RTD. The green dotted line still marks the $43.88$~s minimum OTD, and the red dashed line corresponds to this minimum OTD rescaled as a function of the redshift of the source: $43.88/(1+z)$~s. Many sources, which were observed by Swift/XRT with an OTD greater than $43.88$~s, would not have been deemed interesting from the early X-ray emission point of view. However, thanks to their large cosmological redshift, when looking at their RTD, it is clear that they have been observed $10$~s after the trigger and allow us to observe the new physical process in Episode~II related to the $\nu$NS-rise of GRBs (see Appendix~\ref{app:A}, for a summary of the BdHN model and the emission Episodes).

\begin{figure}
\centering
\includegraphics[width=\hsize]{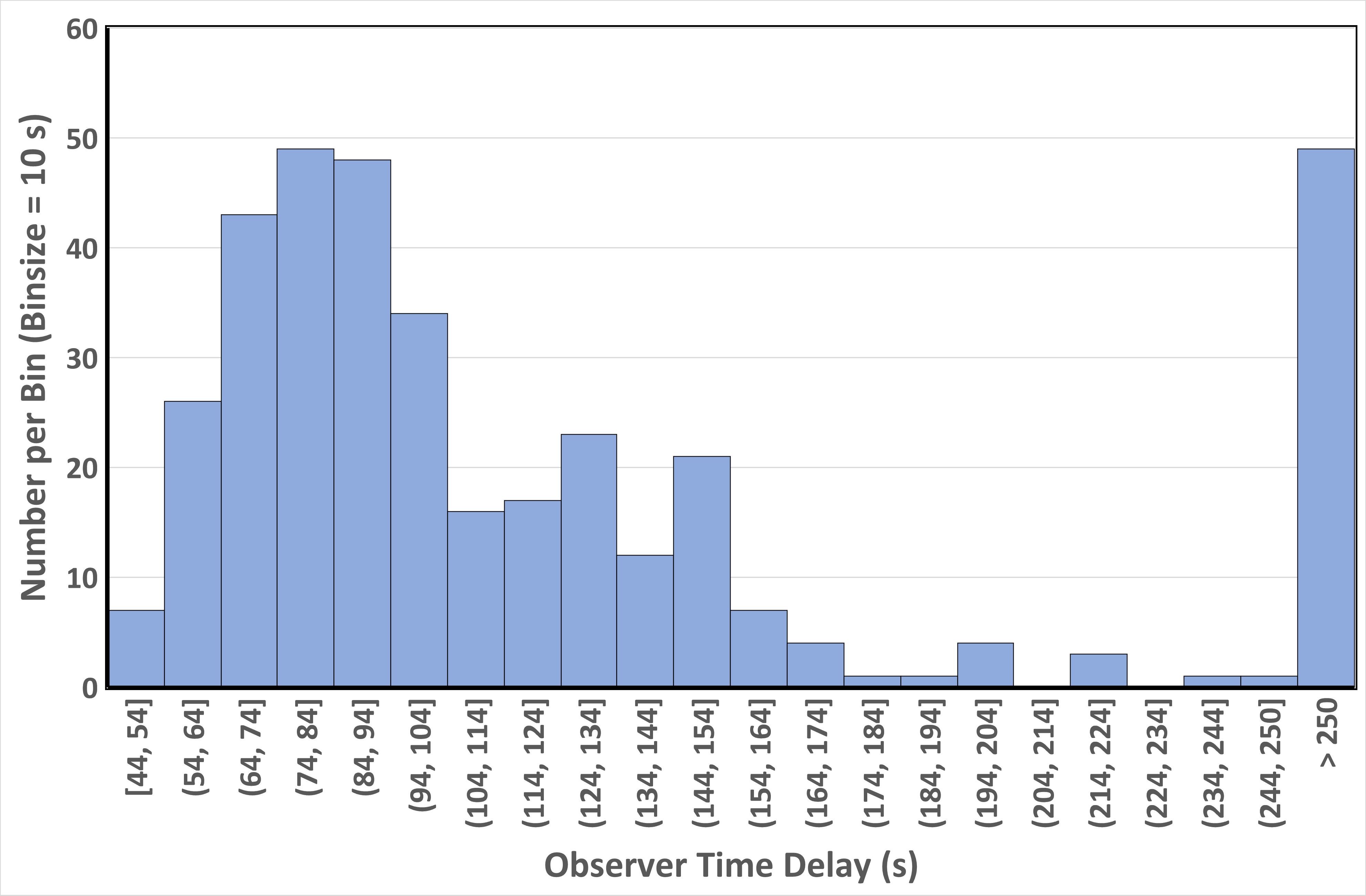}\\
\includegraphics[width=\hsize]{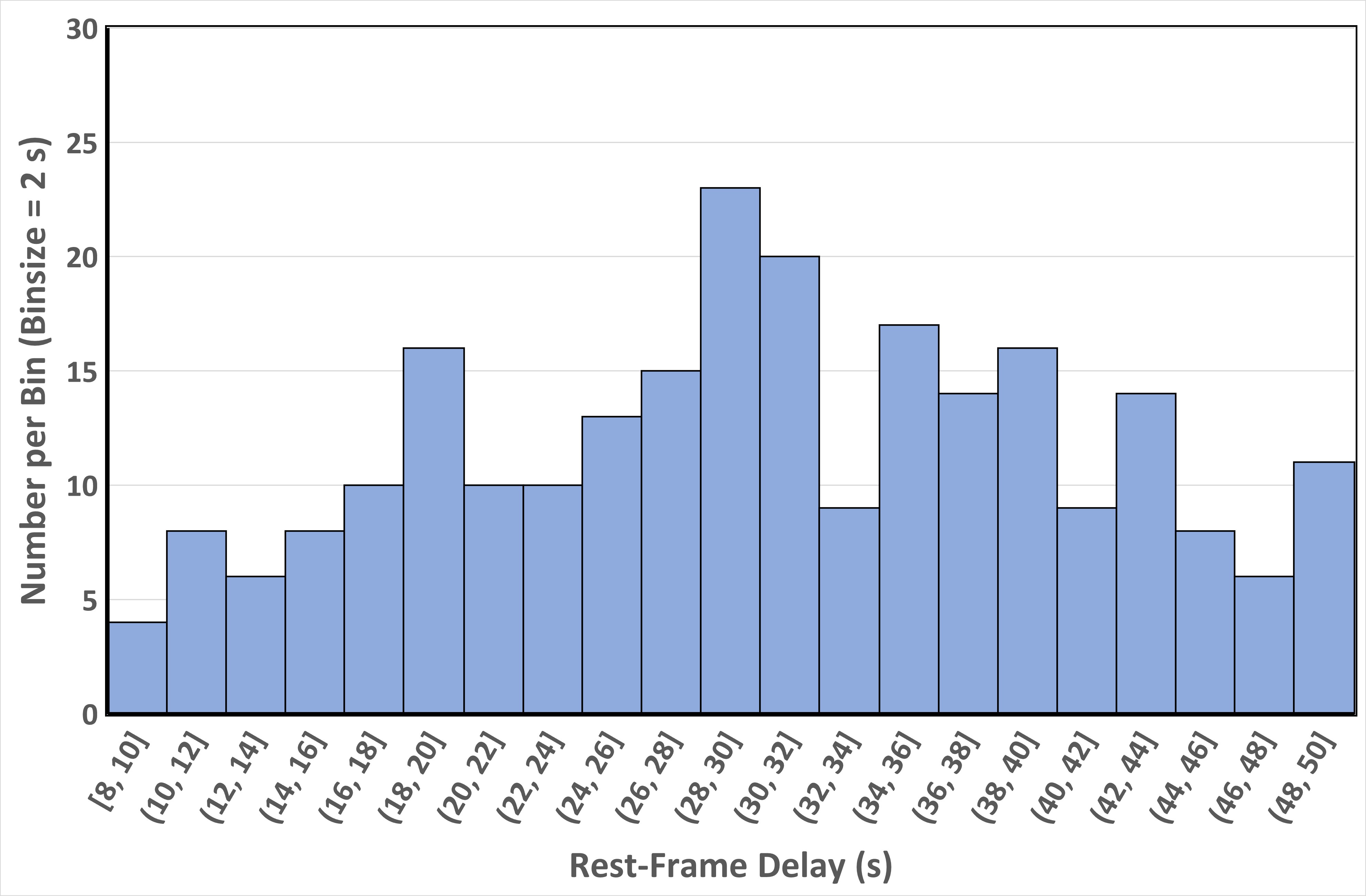}
\caption{The histogram of Swift-XRT time delays in the observer's frame (OTD, upper panel) and the cosmological rest-frame (RTD, lower panel).}\label{fig:histgram}
\end{figure}

This can also be seen in Fig.~\ref{fig:histgram}, where we plot the histogram of the OTD (upper panel) and of the RTD (lower panel): the OTD for most GRBs lies between $50$~s and $170$~s and peaks at $\sim 80$~s, while the recorded minimum RTD in the sample of $368$~GRBs is $\sim 8$~s from GRB~090423 at redshift $z=8.2$ and the RTD range for most of the GRBs is between $8$~s and $50$~s, with a peak at $\sim 30$~s.

Therefore, observing GRBs at large $z$ represents an invaluable tool for exploring the early transient X-ray regimes that occur just after the GRB trigger time and poses a stringent test for all GRB theoretical models. This is particularly relevant in the case of the BdHN model (see Appendix~\ref{app:A}) since, analyzing the transient X-ray regimes occurring just after the SN-rise described in Episode~I and the GRB trigger, it is possible to unveil the physical processes taking place during the $\nu$NS-rise (Episode~II).

\section{The prototypical cases of GRB~220101A, GRB~090423, GRB~090429B}\label{sec:protot}

We now turn to the X-ray emission of three high-$z$ BdHNe~I, our prototypical cases: GRB~220101A, GRB~090423, and GRB~090429B. The photon index during the early afterglow of a GRB exhibits significant variations, especially in the steep decay or X-ray flare periods, where the photon index can deviate from the average value of $\sim2$ in the afterglow, evolving between approximately $1$ and $4$. When calculating the GRB luminosity based on the observed flux, we need to consider the $k$-correction, a function of the photon index. Therefore, we must consider time-resolved $k$-correction when dealing with early afterglow data. For some bursts, the shape of the luminosity light curve of the early afterglow generated by time-resolved $k$-correction differs from that generated by time-integrated $k$-correction \citep[see details in][]{2018ApJ...852...53R, 2023ApJ...945...95W}.

\begin{figure}
\centering
\includegraphics[width=\hsize]{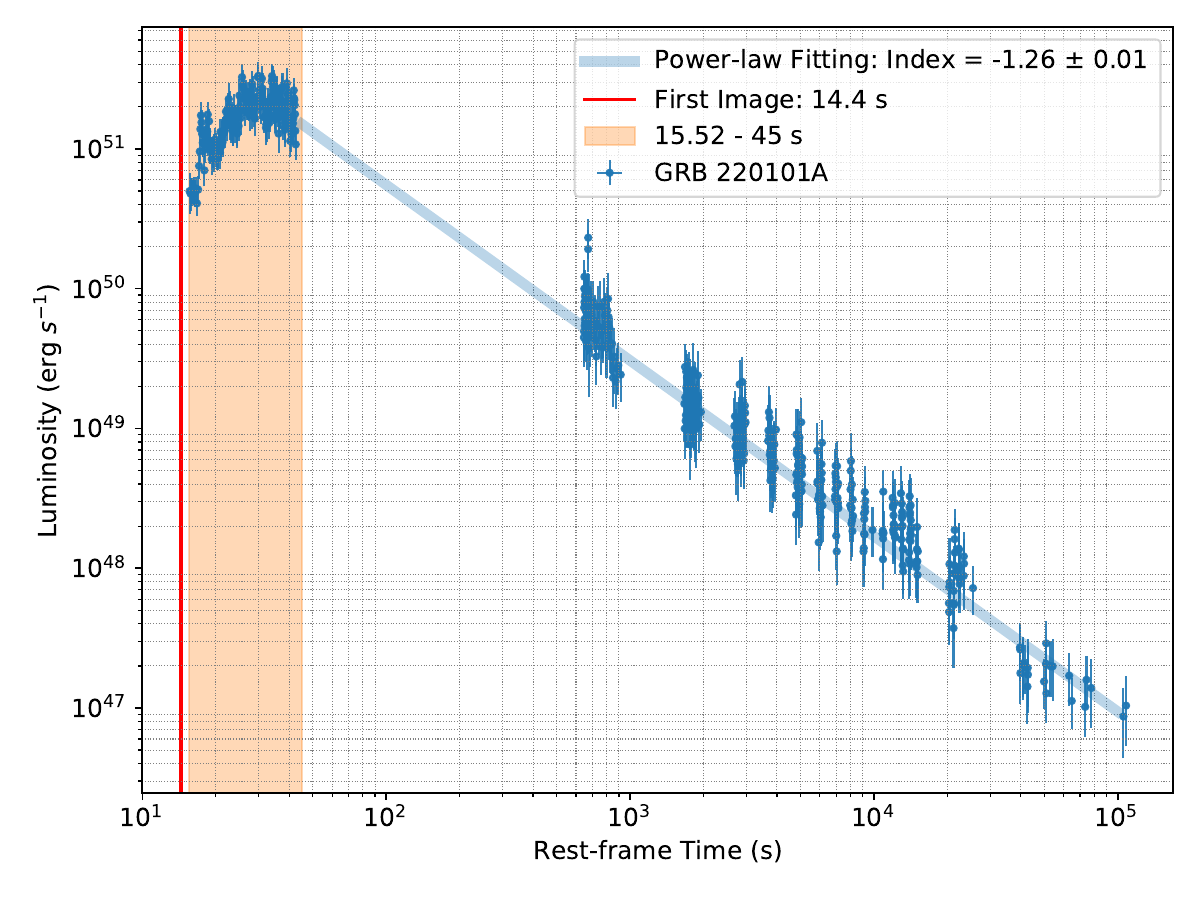}
\includegraphics[width=\hsize]{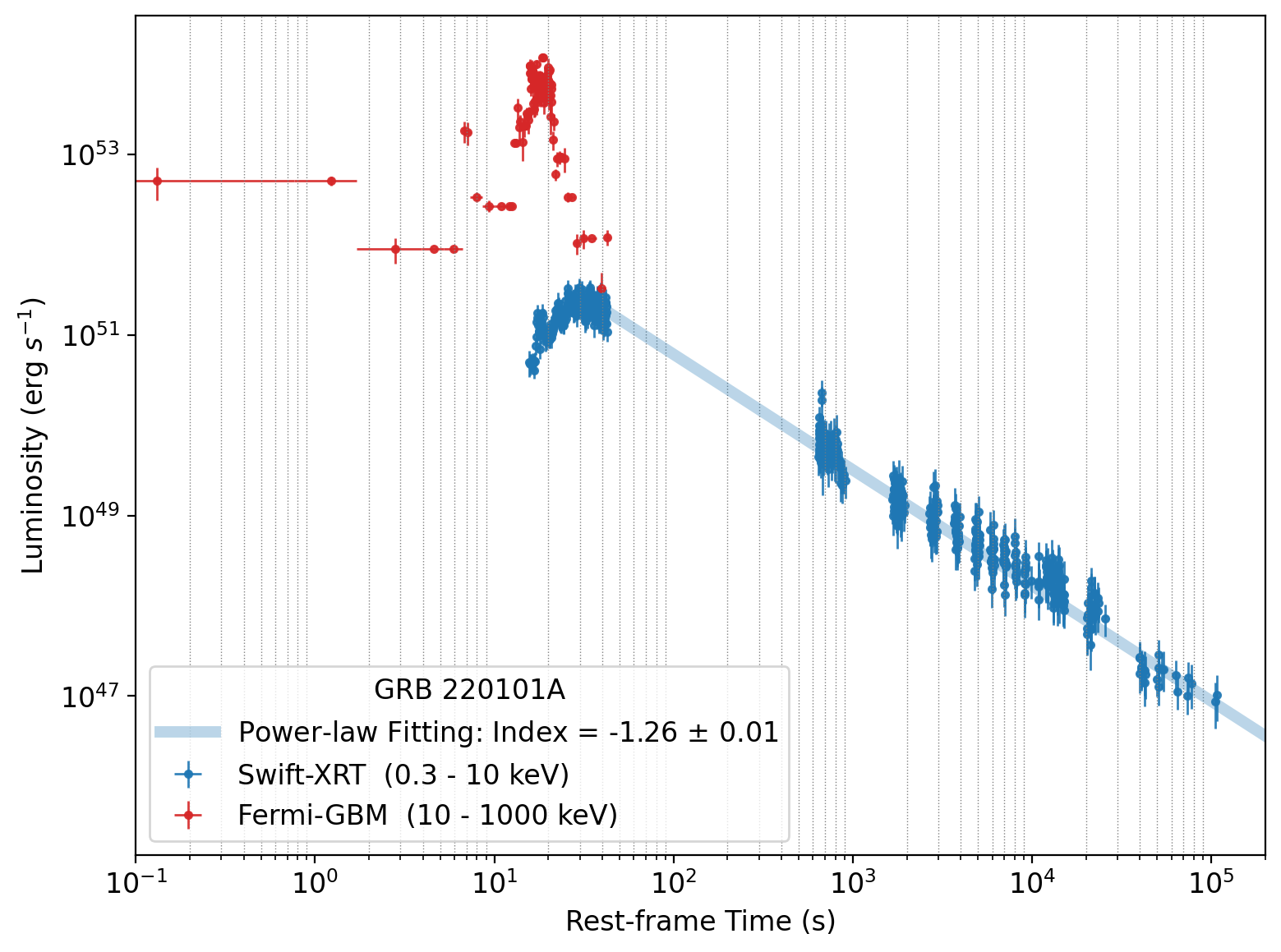}
\caption{\textbf{Upper panel:} The Swift-XRT $0.3$--$10$~keV luminosity of GRB~220101A in the cosmological rest-frame. The red line at $14.4$~s corresponds to the first observation by XRT while still in Image mode before switching to Windowed Timing (WT) mode (for details, see, e.g., E. Troja, ``The Neil Gehrels Swift Observatory Technical Handbook Version 17.0'', \url{https://swift.gsfc.nasa.gov/proposals/tech_appd/swiftta_v17.pdf}, as well as \citealp{2004ApJ...611.1005G}). The orange strip, which extends from $15.52$~s to $45$~s, indicates the data observable thanks to the cosmological effect at $z = 4.61$ duly considered in this article. Other data points between $13.3$~s and $14.4$~s correspond to observations performed while Swift was still slewing to the source location and have not been considered in this paper. The blue line is a power-law fitting function of the form $A_Xt^\alpha$ whose best-fit parameters are: $A_{\rm X}= (1.80 \pm 0.11) \times 10^{53}$~erg s$^{-1}$, and $\alpha = -1.26 \pm0.01$. \textbf{Lower panel:} The same Swift-XRT light curve of the upper plot together with the Fermi-GBM $10$--$10^3$~keV light curve indicating CO core and the SN-rise (Episode~I). This process occurs between $-0.18$~s up to $3.57$~s and lasts for $3.57$~s all in the rest frame. The total energy emitted in this event is $1.2 \times 10^{53}$~erg. The corresponding data for seven additional SN-rise events are now available (Ruffini, Liang Li, and Wang Yu, to be submitted).}\label{fig:delay3}
\end{figure}

\begin{figure}
\centering
\includegraphics[width=\hsize]{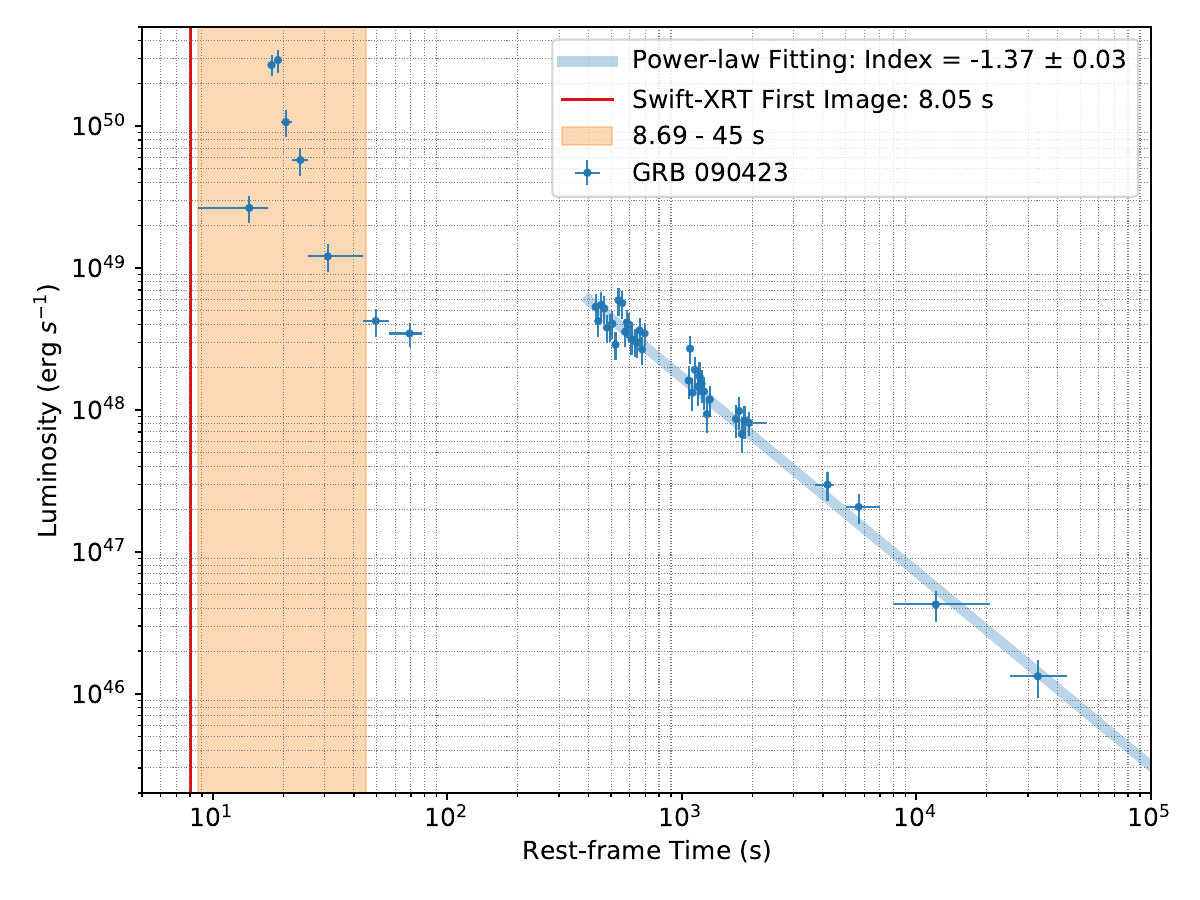}
\includegraphics[width=\hsize]{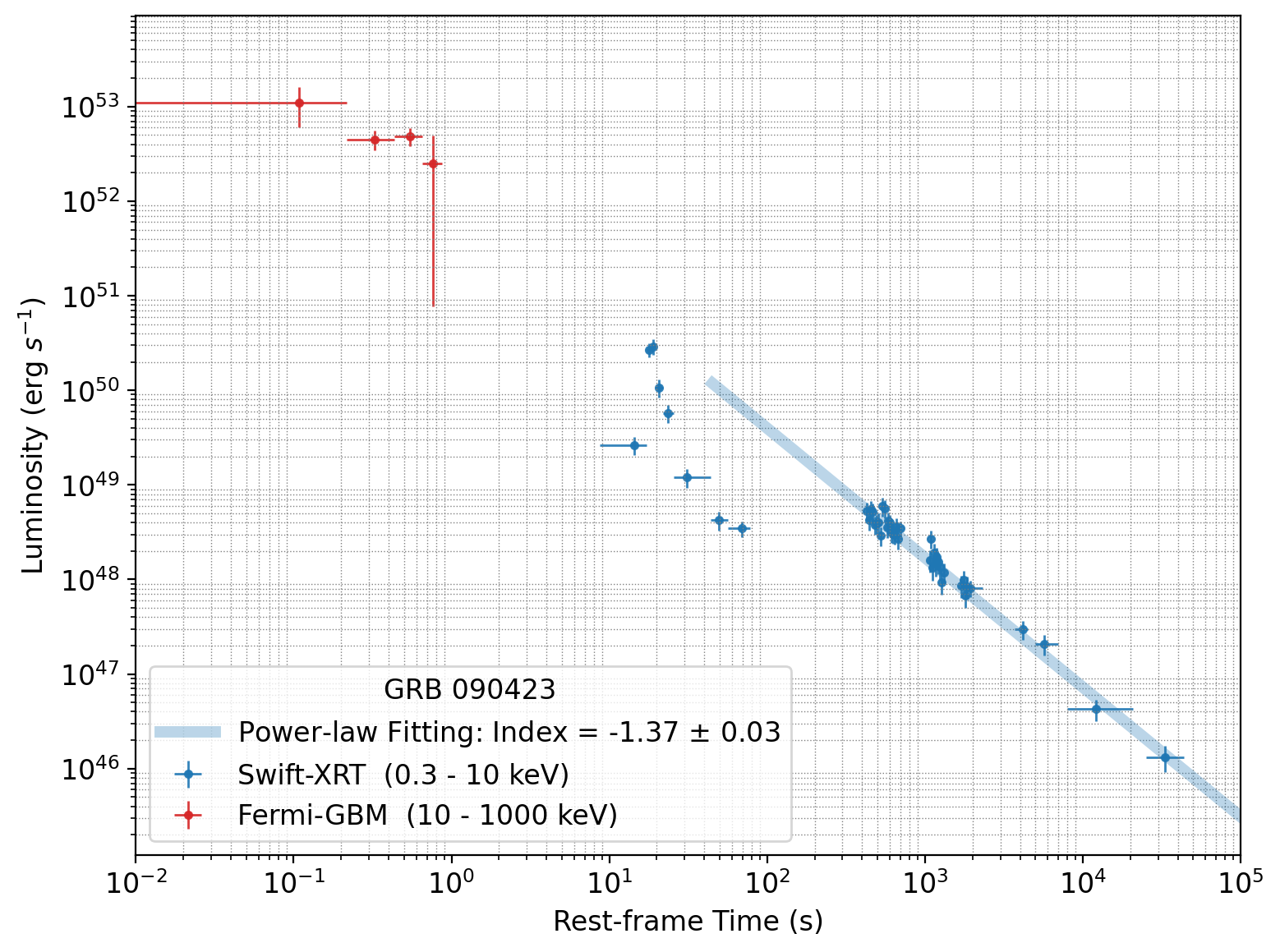}
\caption{\textbf{Upper panel:} The Swift-XRT $0.3$--$10$~keV luminosity of GRB~090423 in the cosmological rest-frame. Same as Fig.~\ref{fig:delay3}, the red line corresponds to the first observation by XRT while still in Image mode before switching to Windowed Timing (WT) mode. The orange strip, which extends from $8.69$~s to $45$~s, indicates the data observable due to the cosmological effect at $z = 8.2$. The blue line is a power-law fitting function of the form $A_Xt^\alpha$ whose best-fit parameters are: $A_{\rm X}= (2.18 \pm 0.49) \times 10^{52}$~erg s$^{-1}$, and $\alpha = -1.37 \pm0.03$. \textbf{Lower panel:} The same Swift-XRT light curve of the upper plot and the Fermi-GBM $10$--$10^3$~keV light curve indicating CO core and the SN-rise (Episode~I). This process occurs between $-0.6$~s to $0.8$~s with a duration of $1.4$~s all in the rest frame. The total energy emitted in this event is $1.6 \times 10^{53}$~erg.}\label{fig:delay4}
\end{figure}

\begin{figure}
\centering
\includegraphics[width=\hsize]{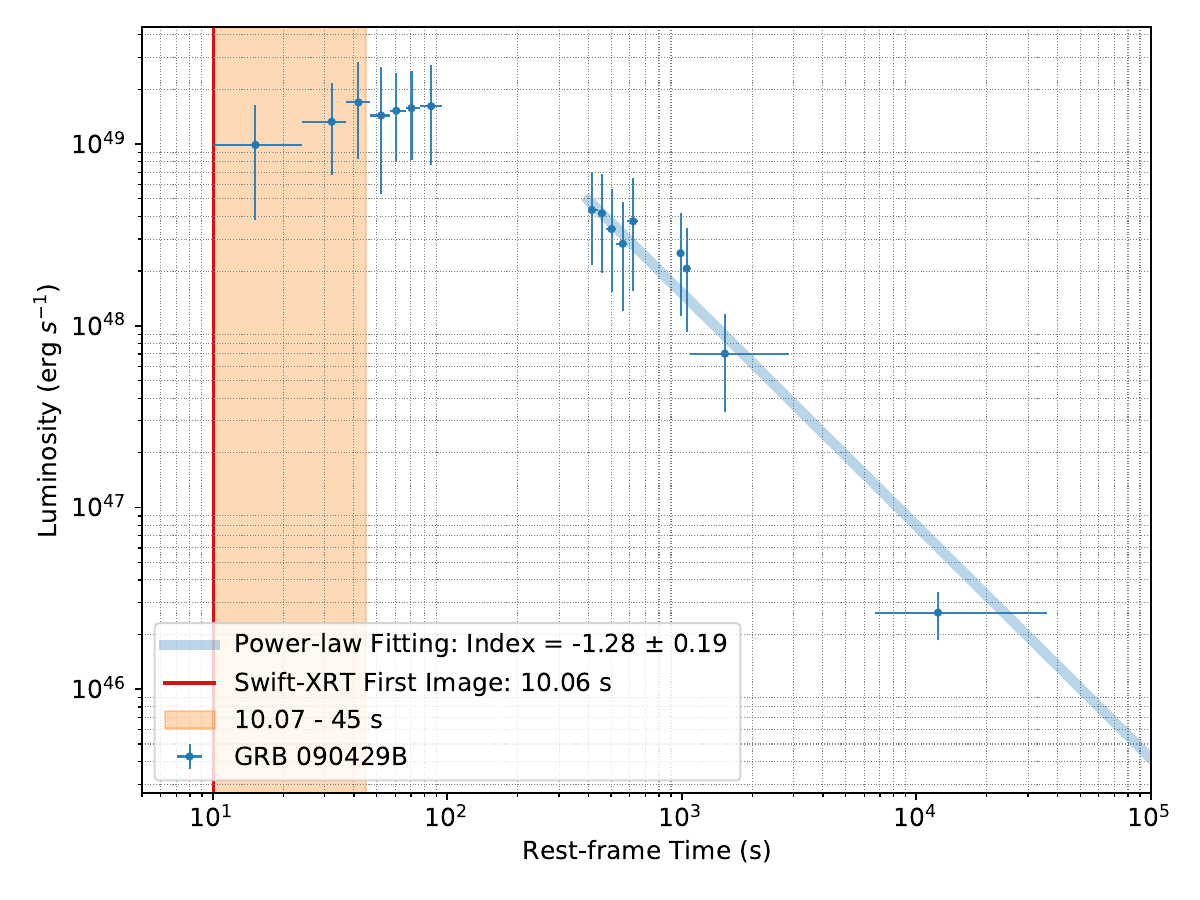}
\includegraphics[width=\hsize]{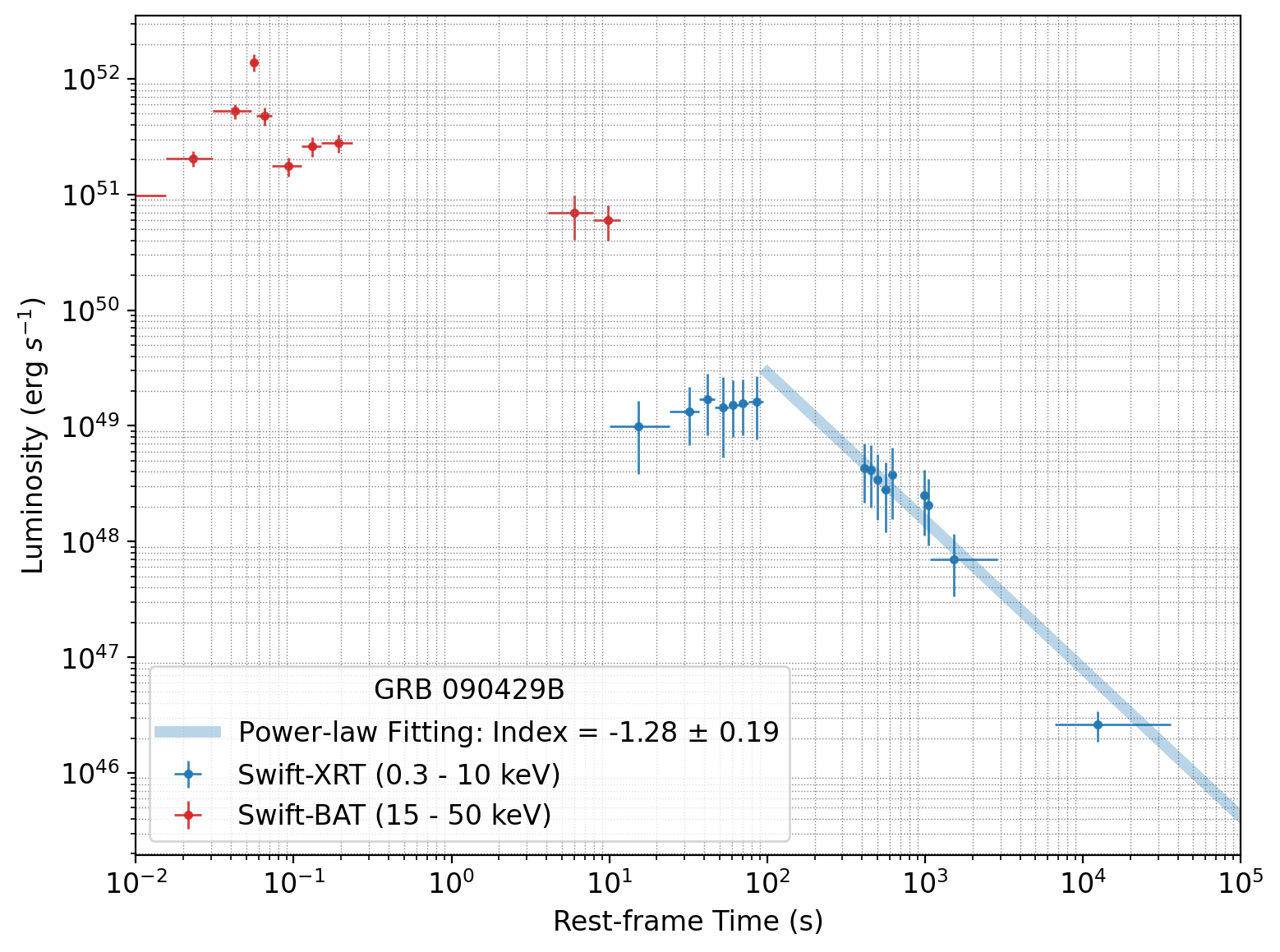}
\caption{\textbf{Upper panel:} The Swift-XRT $0.3$--$10$~keV luminosity of GRB~090429B in the cosmological rest-frame. The red line corresponds to the first observation by XRT while still in Image mode before switching to Windowed Timing (WT) mode. The orange strip, which extends from $10.07$~s to $45$~s, indicates the data observable due to the cosmological effect at $z = 9.4$. The blue line is a power-law fitting function of the form $A_Xt^\alpha$ whose best-fit parameters are: $A_{\rm X} = (1.05 \pm 0.13) \times 10^{52}$~erg s$^{-1}$, and $\alpha = -1.28 \pm0.19$. \textbf{Lower panel:} The same Swift-XRT light curve of the upper plot and the Swift-BAT $15$--$50$~keV light curve indicating the CO core and the SN-rise (Episode~I). This process occurs between $0$~s and $0.96$~s with a duration of $0.96$~s in the rest frame. The total energy emitted in this event is $3.5 \times 10^{52}$~erg.}\label{fig:delay5}
\end{figure}

From the above analysis, we conclude for the prototypical cases:
\begin{itemize}
    \item GRB~220101A has a redshift $z=4.61$, the OTD is $80.78$~s corresponding to an RTD of $14.40$~s. Swift/XRT $0.3$--$10$~keV luminosity is shown in Fig.~\ref{fig:delay3}. The orange strip marks the data before $\sim 45$~s, which are observable only thanks to the methodology presented in this article for high source redshift. The best-fit parameters of the decaying part are $A\rm_X= (1.80 \pm 0.11) \times 10^{53}$~erg s$^{-1}$, and $\alpha = -1.26 \pm 0.01$ representing the X-ray afterglow.
    \item GRB~090423 has a redshift $z\sim8.2$, the OTD is $72.48$~s corresponding to an RTD of $\sim 8$~s. Swift-XRT $0.3$--$10$~keV luminosity is shown in Fig.~\ref{fig:delay4}. Same as Fig.~\ref{fig:delay3}, the orange strip marks the data from $8.1$~s up to $\sim45$~s. The best-fit parameters of the decaying part are $A\rm_X= (2.18 \pm 0.49) \times 10^{52}$~erg s$^{-1}$, and $\alpha = -1.37 \pm0.03$ representing the X-ray afterglow.
    \item GRB~090429B has a photometric redshift $z\sim9.4$. The OTD is $104.69$~s, corresponding to an RTD of $\sim 10.1$~s. Swift-XRT $0.3$--$10$~keV luminosity is shown in Fig.~\ref{fig:delay5}. The orange strip marks the data between $10.1$~s to $\sim 45$~s. The best-fit parameters of the decaying part are $A\rm_X= (1.05 \pm 0.13) \times 10^{52}$~erg s$^{-1}$, and $\alpha = -1.28 \pm0.19$ representing the X-ray afterglow.
\end{itemize}

Therefore, the above analysis of the X-ray emission observed by Swift/XRT is a powerful tool to validate the observation of the Episodes expected in the BdHN scenario (see Appendix~\ref{app:A}). For the three prototypes analyzed above, the XRT data shows, at the end of the SN-rise (Episode~I), the presence of Episode~II marked by the spin-up phase of the $\nu$NS by SN ejecta fallback accretion \citep{2022PhRvD.106h3004R}, followed by its slowing down characterizing the X-ray afterglow \citep{2023ApJ...945...95W,2022ApJ...939...62R,2022PhRvD.106h3002B,2022PhRvD.106h3004R}. These results confirm the sources' BdHN I nature and the formation of stellar-mass BHs up to large cosmological redshifts, $z\sim 10$.

This analysis will be completed by information on emission in all the wavelengths in the GeV and MeV for redshifts smaller than $5$, e.g., GRB~220101A (R. Ruffini, Wang Yu., et al., to be submitted), see Fig~\ref{fig:delay3}.

\section{Inference for the cosmological distribution of the black holes}\label{sec:6}

The analysis of GRB~220101A, GRB~090423, GRB~090429B heralds an important astrophysical message. We have shown that stellar-mass BHs from BdHNe~I occur at a very high redshift, $z\sim 10$, originating from massive binary stars, possibly $\lesssim 25 M_\odot$ each, only a few hundred million years from the Big Bang. This conclusion may suggest revisiting the cosmological and stellar evolution paradigm. Indeed, our results agree with the conclusion by \citet{2014ARA&A..52..415M}: ``it seems premature to tinker further with the (stellar initial mass function) IMF, although if discrepancies remain after further improvements in the measurements and modeling then this topic may be worth revisiting.''

The above evidence of stellar-mass BHs formed in GRBs up to redshift $z\sim 10$ complements the daily information being gained by the observations of the James Webb Space Telescope (JWST) of quasars at high redshift, e.g., at $z\sim 6$--$7$ \citep{2021ApJ...923..262Y,2023arXiv230904614Y}, and the farthest quasars ever observed at the center of the galaxy UHZ1 at $z\approx 10.3$  \citep{2023NatAs.tmp..223B} and of GN-z11 at $z\approx 10.6$ \citep{maiolino2024}. These observations are unveiling a larger population of supermassive BHs at very high redshift than previously thought (see  \citealp{2022A&A...666A..17G,2023arXiv230801230M,2023ApJ...954L...4K} and \citealp{2023ARA&A..61..373F} for a recent review), suggesting a possible role of dark matter in their formation \citep{2023MNRAS.523.2209A, 2024ApJ...961L..10A}. Therefore, from all the above, it appears that the presence of BHs in the Universe is ubiquitous. Indeed, these two topics are not independent. A new research window is open to test whether supermassive BHs at large redshifts may boost the star formation in the early Universe \citep{2022NewAR..9401642M}.

\section{The BdHN model and the double-peak GRB redshift distribution}\label{sec:subsample}

\begin{figure}
\centering
\includegraphics[width=\hsize]{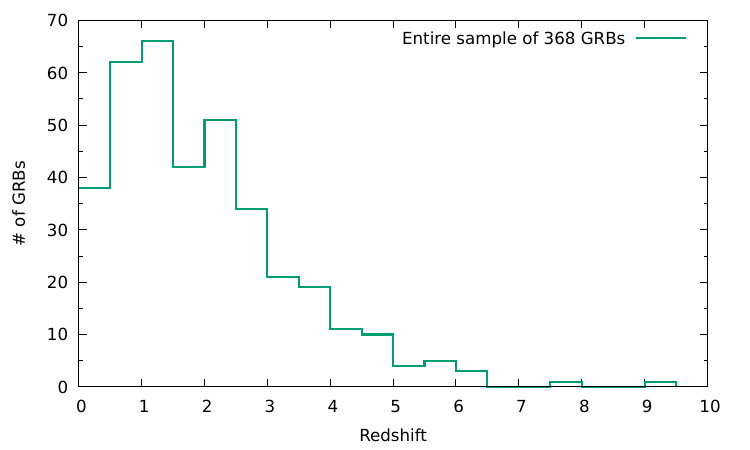}
\caption{The distribution of the redshifts of the $368$ GRBs in our sample (see Table~\ref{tab:list-XRT})}\label{fig:z_dist_354}
\end{figure}

We turn now to additional information that can be gained from the distribution of GRBs across the Universe. In Fig.~\ref{fig:z_dist_354}, we present the distribution of the redshifts of the $368$~GRBs in our sample (see Table~\ref{tab:list-XRT}). We can see the first peak between $z = 1$ and $z=1.5$ and the second peak between $z=2$ and $z=2.5$. This same \textit{double-peak} structure in the GRB redshift distribution was also present in the GRB sample considered by \citet{2010MNRAS.406.1944W} and in the one considered by \citet{2016ApJ...829....7L}.

Having determined the redshift distribution of our sample of $368$~GRBs detected by Swift since the year 2005 up to the end of the year 2023, including both long and short GRBs, we would like to address the double peak structure found in Fig.~\ref{fig:z_dist_354}. We are going to use all available data on different BdHN families.
 
As indicated in Appendix~\ref{app:A}, see Table~\ref{tab:episodes}, the BdHN model identifies several different GRB families \citep[][and references therein]{2023ApJ...955...93A}, and it is of interest to inquire about the possible difference in the redshift distribution of each family.

\begin{figure}
\centering
\includegraphics[width=\hsize]{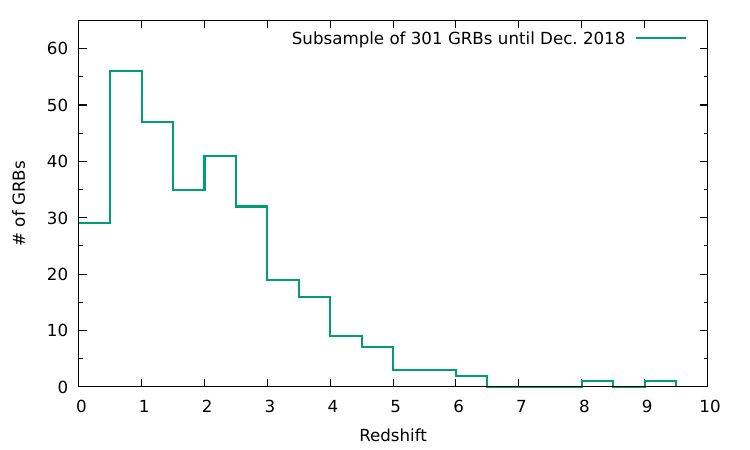}
\caption{The distribution of the redshifts of the $301$ GRBs in the subsample until Dec. 2018.}\label{fig:z_dist_301}
\end{figure}

A preliminary result was obtained by \citet{2021MNRAS.504.5301R} where a catalog of all BdHNe~I observed from the early 1990s until December 2018 is presented. Therefore, we limit our current analysis of the redshift distribution of our $368$~GRB sample to a subsample of $301$ GRBs exploded until December 2018 (see below and Fig.~\ref{fig:z_dist_301}). What is very interesting is that also, in this distribution, the double peak structure is maintained. We can then compare and contrast this redshift distribution of Fig.~\ref{fig:z_dist_301} with the BdHN~I sample published in \citet{2021MNRAS.504.5301R}. 
\begin{itemize}
    \item We build a subsample of our $368$~GRB sample by selecting all GRBs detected until December 2018. There are $304$~GRBs in this subsample.
    \item We look for each of these $304$~GRBs in the BdHNe~I catalog published by \citet{2021MNRAS.504.5301R}. We find $216$~of them. We, therefore, conclude that in our subsample of $304$~GRBs exploded until December 2018, there are $216$~BdHNe~I.
    \item We have $88$~GRBs in our subsample that still need to be classified.
    \item We look to the observed prompt emission duration of each $88$~GRBs still needing a classification.
    \item We see that $21$~GRBs have an observed prompt emission duration $T_{90}<2$~s and can therefore be classified as short GRBs. They are too few to be further subdivided to build statistically significant redshift distributions of the different families of short GRBs implied by the BdHN model. Therefore, our current analysis considers the cumulative redshift distribution of all these $21$~short GRBs as a single family.
    \item We are left with $67$~GRBs in our subsample that still need a classification.
    \item We find that $64$~GRBs have an observed prompt emission duration $T_{90}>2$~s. Therefore, they are neither short GRBs nor BdHNe~I \citep[since they are not in the catalog by][albeit exploded before December 2018]{2021MNRAS.504.5301R}. They must be either BdHNe~II or BdHNe~III. A further subdivision of these $64$~GRBs into BdHNe~II and BdHNe~III requires an extra analysis outside the present paper's scope. Then, in our current analysis, we are considering the cumulative redshift distribution of all these $64$~BdHNe~II or BdHNe~III as if they were a single family.
    \item We still have $3$~GRBs in our sample with no observed $T_{90}$ duration in the Swift catalog, which we therefore exclude from our current analysis. Our final subsample of GRBs exploded until December 2018 and contains $301$ GRBs.
\end{itemize}
In summary, starting from our sample of $368$ GRBs, we built a subsample of $301$~GRBs detected until December 2018. The redshift distribution of the GRBs in this subsample is plotted in Fig.~\ref{fig:z_dist_301} to be compared and contrasted with the one of the entire sample of $368$~GRBs (Fig.~\ref{fig:z_dist_354}). We can see that both distributions present the same double-peaked structure at $z\lesssim 2$. Therefore, the results we will obtain by analyzing the distribution of the subsample can be considered valid for the entire sample of $368$~GRBs as well.

\begin{figure}
\centering
\includegraphics[width=\hsize]{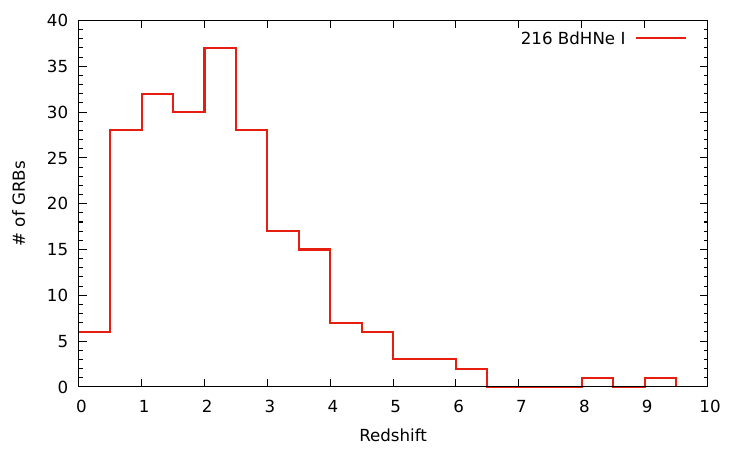}
\includegraphics[width=\hsize]{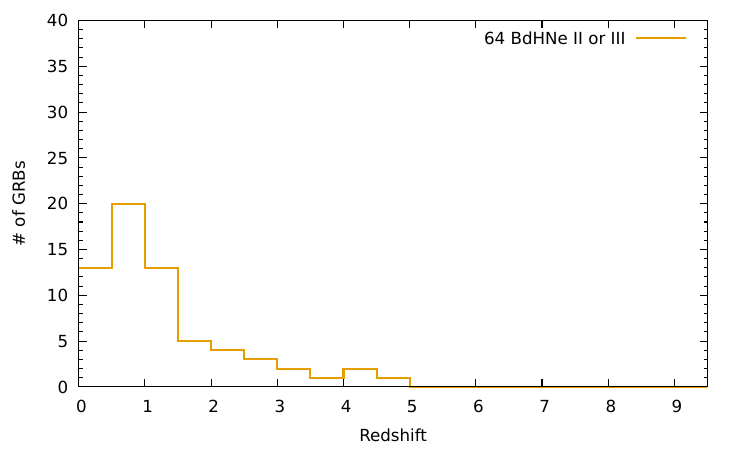}
\includegraphics[width=\hsize]{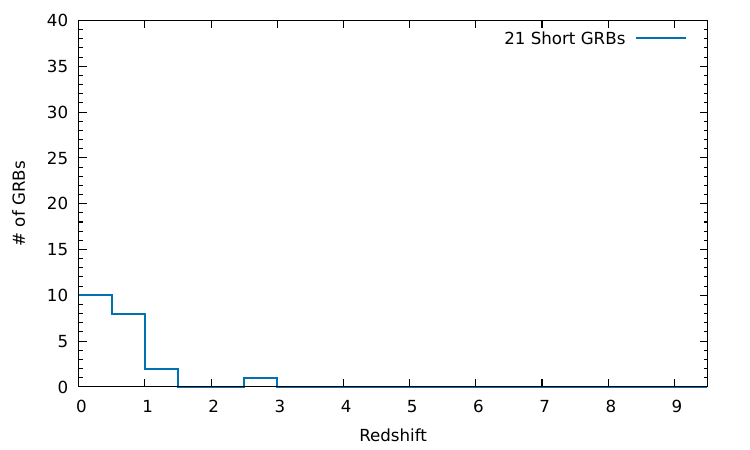}
\caption{The distributions of the redshifts of the $301$~GRBs in our subsample, divided into three groups: BdHNe~I (upper panel, red line), BdHNe~II/III (middle panel, orange line), Short GRBs (lower panel, blue line).}\label{fig:z_dist_split}
\end{figure}

Following the procedure described above, we have that the subsample of $301$~GRBs exploded until December 2018 can be subdivided among the different GRB families indicated by the BdHN model as follows:
\begin{itemize}
    \item $216$ GRBs are BdHNe~I;
    \item $64$ GRBs are BdHNe~II or BdHNe~III;
    \item $21$ GRBs are short GRBs.
\end{itemize}
Fig.~\ref{fig:z_dist_split} shows the distributions of the redshifts of each of these three GRB groups in the subsample. We can see that the redshift distribution of BdHNe~I presents a single peak between $z\sim 2$ and $z\sim 2.5$ and a sort of plateau for $0.5 \lesssim z \lesssim 2$, while the distribution of BdHNe~II and BdHNe~III presents a single peak around $z\sim 1$ and that of short GRBs presents a single peak for $z<0.5$. The K-S test applied to the distributions shown in Fig.~\ref{fig:z_dist_split}, BdHNe~I (top panel) vs. Short (bottom panel), yields a probability $P=4.5\times10^{-10}$ suggesting that there is sufficient statistical evidence to conclude that the redshift distributions of BdHNe~I and short GRBs are not identical. The same conclusion can be reached after comparing BdHNe~I (top panel) vs. BdHNe~II and BdHNe~III (middle panel), which yields a $P=5.0\times10^{-9}$. On the other hand, the K-S test applied to the redshift distributions of BdHNe~II and BdHNe~III (middle panel) vs. short GRBs (bottom panel) yields a much larger value, $P=0.011$, indicating that the two distributions do not differ significantly. This similarity in the redshift distributions supports the idea, advanced in \citet{2016ApJ...832..136R,2018ApJ...859...30R}, that BdHNe~II and BdHNe~III may end up in remnant binary systems that, in turn, at the end of their evolution, can later become progenitors of short GRBs. We can also see that BdHNe~I extend to much higher redshifts than BdHNe~II, BdHNe~III, and short GRBs, but this may be due to a selection effect (BdHNe~I, being the most energetic and most luminous long GRBs, are easier to detect even at very high redshift).

We can then conclude that the double peak structure in the redshift distribution of our sample of $368$~GRBs can be explained by the superposition of the redshift distributions of the different BdHN families.
\section{Conclusion}\label{sec:7}

We can summarize three main conclusions:
\begin{enumerate}
\item 
In this article, we have introduced the use of the time dilation in high-redshift GRBs to overcome the observed instrumental time delay, greater than $43$~s, between the GRB trigger time and the first X-ray observations by Swift/XRT. This time delay has traditionally hampered the observations of Episode~I and Episode~II in BdHNe \citep[see, e.g.,][]{2023ApJ...955...93A}. The methodology has been developed using a sample of $368$ GRBs, reported in Table~\ref{tab:list-XRT}, all with an identified redshift. When measured in the observer frame, the time delay (OTD) between the earliest X-ray emission and the GRB trigger time is always larger than $40$~s (see the upper panels of Fig.~\ref{fig:delay1} and of Fig.~\ref{fig:histgram}). In contrast, a substantially shorter time delay is observed in the rest frame of the source (RTD, see the lower panels of Fig.~\ref{fig:delay1} and of Fig.~\ref{fig:histgram}). This new methodology allows the analysis of the very early transient X-ray regimes in GRB afterglows, which pose a stringent test for all GRB theoretical models. Within the context of the BdHN model, we applied it to three BdHNe~I at high redshift. This has allowed us to unveil the occurrence of the spinning up $\nu$NS emission, increasing with time. This emission precedes the traditional X-ray afterglow emission, which decreases in time with a specific power-law index.
\item 
The most eloquent example is the case of one of the most powerful GRBs ever detected, GRB~220101A, at $z=4.61$ (Ruffini et al. 2024 to be submitted). Given the source's high redshift and outstanding luminosity this source allows the identification of all seven episodes of a BdHN, except for the late radioactive decay of the SN ashes. In particular, GRB~220101A shows the SN-rise (Episode~I) triggering the entire GRB, see Fig~\ref{fig:delay3}  (Ruffini et al., 2023, in preparation). Especially significant are the unexpected high-quality data associated with the Swift/XRT observations of the $\nu$NS-rise (Episode~II). The X-ray emission observed by Swift/XRT starts $14.4$~s after the trigger, following the end of the SN-rise and indicating the spin-up phase of the $\nu$NS by the fallback accretion of matter initially ejected by the SN. It is followed by the slowing down phase starting at $45$ s corresponding to the decaying part of the X-ray afterglow \citep{2023ApJ...945...95W,2022ApJ...939...62R,2022PhRvD.106h3002B,2022PhRvD.106h3004R}. The unexpected very high-quality data associated with the Swift/XRT observations of the $\nu$NS-rise (Episode~II) also applies to GRB~090423 at $z=8.2$ (see Fig.~\ref{fig:delay4}) and GRB~090429B at $z=9.4$ (see Fig.~\ref{fig:delay5}): in both cases the first Swift/XRT data shows the $\nu$NS spin-up phase, extending up to $10^2$~s and followed by the slowing down phase corresponding to the decaying part of the X-ray afterglow. One of the key questions to be addressed is the possibility that, at the end of the spin-up phase, a short time ($\lesssim 1$ s) process of gravitational wave emission occurs due to a transition to a triaxial configuration of the fast spinning $\nu$NS, with characteristic strain $h_c\sim 10^{-23}$ at about kHz frequency \citep[see][for details]{2022PhRvD.106h3004R}.
\item 
Equally important is the byproduct of analyzing the redshift distributions of all the $368$ GRBs of the sample (see Fig.~\ref{fig:z_dist_354}), of all the $301$ GRBs until December 2018 (see Fig.~\ref{fig:z_dist_301}), and in particular of the $216$ BdHNe~I, of the $64$ BdHNe~II and BdHNe~III, and of the $21$ short GRBs until December 2018 (see Fig.~\ref{fig:z_dist_split}). The distribution of the entire sample of $368$ sources presents two peaks: the first, dominated by the BdHNe~II and BdHNe~III, at $z\sim 1$ and the second, dominated by BdHNe~I, at $z \sim 2$. Such a two-peak structure of the GRB density rate, which seems not to trace the star formation rate \citep[see, e.g.,][]{2014ARA&A..52..415M}, is indeed present in the distributions by \citet{2010MNRAS.406.1944W} and \citet{2016ApJ...829....7L}. We have shown here that the different physical properties between BdHNe~I, BdHNe~II and BdHNe~III explains these two peak distribution. An additional conclusion can be drawn based on the GRB distributions. The similarity between the redshift distribution of BdHNe~II and BdHNe~III and that of short GRBs supports the hypothesis, advanced in \citet{2016ApJ...832..136R,2018ApJ...859...30R}, that the BdHNe~II and BdHNe~III remnants, after evolving into binary NS systems, could later become progenitors of short GRBs. This unique prediction of the BdHN scenario deserves further attention from an observational and a theoretical point of view, e.g., recent simulations show BdHNe lead to bound NS-NS binaries with a wide range of merger times \citep{2024arXiv240115702B}.
\end{enumerate}

Indeed, a great opportunity exists for new missions with wide field-of-view soft X-ray instruments designed to simultaneously observe the GRB X-ray and gamma-ray emissions from $0.3$~keV to $10$~MeV from the moment of the GRB trigger without any time delay, such as, e.g., THESEUS \citep{2018AdSpR..62..191A,2021ExA....52..183A} and HERMES \citep{2019NIMPA.936..199F,2020SPIE11444E..1RF}.

\acknowledgments
We express our gratitude to an anonymous referee whose suggestions greatly improved the presentation of our results. We are grateful to B. Cenko and J. Kennea for their important clarifications on the timing of the Swift automatic slew system. We also thank Y. Aimuratov, L. Amati, L. M. Becerra, C. Cherubini, M. Karlica, P. Madau, and N. Sahakyan for the fruitful discussions leading to these new results.

\appendix

\section{The binary-driven hypernova model}\label{app:A}

Within the BdHN model, long GRBs of different energies have a common progenitor: a binary comprising a CO star and an NS companion \citep{2012ApJ...758L...7R}. The CO star is at the end of its thermonuclear evolution. The collapse of the CO star iron core forms a newborn NS ($\nu$NS) at the center and produces an SN explosion. The latter triggers various physical phenomena whose occurrence and/or entity depend mostly on the system's orbital period \citep{2014ApJ...793L..36F,2015PhRvL.115w1102F}. Numerical, three-dimensional, smoothed-particle-hydrodynamics simulations of the BdHN scenario \citep{2016ApJ...833..107B, 2019ApJ...871...14B, 2022PhRvD.106h3002B} has led to define three different types of BdHNe, corresponding to the diversity of long GRBs:
\begin{enumerate}
    \item BdHNe~I are the most extreme with energies $10^{52}$--$10^{54}$~erg. Their orbital periods are about $5$ minutes. In these sources, the material ejected in the SN is easily accreted by the NS companion, so it reaches the point of gravitational collapse, forming a rotating BH. BdHN~I examples are GRB 130427A \citep{2019ApJ...886...82R}, GRB~180720B \citep{2022ApJ...939...62R}, and GRB 190114C \citep{2021PhRvD.104f3043M, 2021A&A...649A..75M}.
    \item BdHNe~II have orbital periods of $20$--$40$ minutes and emit energies $10^{50}$--$10^{52}$~erg. The accretion is lower, so the NS remains stable. A BdHN~II example is GRB~190829A \citep{2022ApJ...936..190W}.
    \item BdHNe~III have orbital periods of hours, and the accretion is negligible. They explain GRBs with energies lower than $10^{50}$ erg. A BdHN~III example is GRB~171205A \citep{2023ApJ...945...95W}.
\end{enumerate}

\begin{figure}
    \centering
    \includegraphics[width=\hsize,clip]{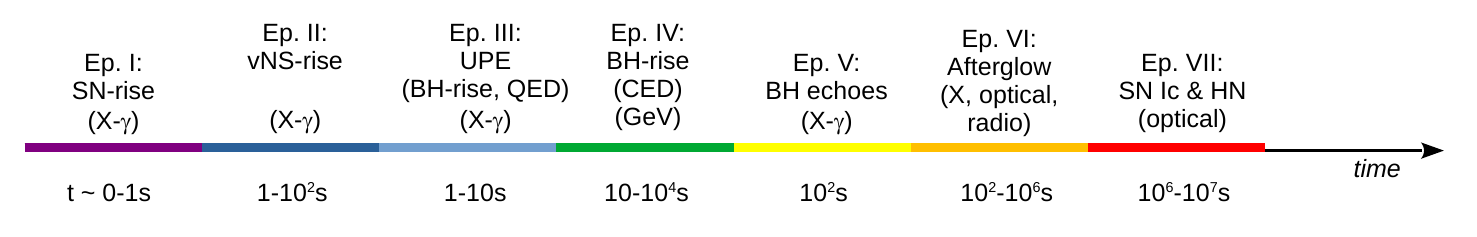}
    \caption{Time sequence of the Episodes in BdHN~I, according to Table~\ref{tab:episodes}. The times are orders of magnitude estimates based on the specific examples of BdHN~I, GRB 180720B \citep{2022EPJC...82..778R, 2022ApJ...939...62R}, and GRB 190114C \citep{2021PhRvD.104f3043M, 2021A&A...649A..75M, 2020ApJ...893..148R}, summarized in \citet{2023ApJ...955...93A}. Some episodes of BdHN~II and III in Table~\ref{tab:episodes} have also been identified, e.g., in GRB~190829A \citep{2022ApJ...936..190W} and GRB~171205A \citep{2023ApJ...945...95W}, as summarized in \citet{2023ApJ...955...93A}. The acronyms are the same as in Table~\ref{tab:episodes}.}
    \label{fig:timeline}
\end{figure}

Seven observable episodes characterize the most general BdHN sequence of physical processes. They have spectral signatures in the GRB precursor, MeV prompt, GeV and TeV emissions, X-optical-radio afterglow, and optical SN emission. They involve the physics of the early SN, NS accretion, black hole (BH) formation, synchro-curvature radiation, and quantum and classic electrodynamics processes to extract the BH rotational energy. Episode~I, the SN-rise, describes the CO core collapse, generating the $\nu$NS and the SN. This episode has been possibly identified in three BdHNe, GRB~090423, GRB~090429B, and GRB~220101A (see Sec.~\ref{sec:protot}). Episode~II, the $\nu$NS- and NS-rise, owes to the SN ejecta accretion onto the $\nu$NS, the NS companion, and their consequent spin-up process \citep{2022PhRvD.106h3002B, 2022PhRvD.106h3004R, 2023ApJ...945...95W}. Episode~III, the ultrarelativistic prompt emission (UPE) phase, is explained by the radiation by an $e^+e^-$ expanding self-accelerated pair plasma loaded with baryons \citep{2021PhRvD.104f3043M, 2022EPJC...82..778R, 2023ApJ...945...10L}. The pairs are produced by the quantum electrodynamics (QED) process of vacuum breakdown by an overcritical electric field. The latter is induced by the interaction of the BH spin with the external magnetic field (inherited by the collapsed NS). Episode~IV, the GeV emission, is due to the radiation of electrons accelerated by the induced electric field near the BH \citep{2019ApJ...886...82R, 2020EPJC...80..300R, 2021A&A...649A..75M, 2022ApJ...929...56R}. Episode~V, called \textit{BH echoes}, encompasses the emission from the interaction of the expanding $e^+e^-$ pairs with the ultralow-density region (referred to as the \textit{cavity}) around the BH \citep{2019ApJ...883..191R} and along other directions of higher density leading to soft and hard X-ray flares (SXFs and HXFs, \citealp{2018ApJ...852...53R}). Episode~VI, the X-ray, optical, and radio afterglow emission, is due to synchrotron radiation in the expanding ejecta and the $\nu$NS pulsar emission \citep{2018ApJ...869..101R, 2019ApJ...874...39W, 2020ApJ...893..148R, 2022ApJ...939...62R}. Finally, Episode~VII is the optical emission by the SN ejecta, powered by the decay of nickel into cobalt \citep[see][for a recent analysis of the SN associated with long GRBs]{2023ApJ...955...93A}. Table~\ref{tab:episodes} and Fig.~\ref{fig:timeline} summarize the physical processes and associated GRB Episodes of each BdHN type, including specific examples; see also \citet{2023ApJ...955...93A} for further details on the BdHN emission episodes. 

\begin{table*}
    \centering
    \caption{Upper: Physical phenomena occurring in BdHN~I, II, and III, and their associated observations in the GRB data. References in the table: $^a$\citet{2019ApJ...874...39W, 2022ApJ...936..190W, 2022PhRvD.106h3004R},$^b$\citet{2014ApJ...793L..36F, 2016ApJ...833..107B, 2022PhRvD.106h3002B, 2022PhRvD.106h3004R, 2022ApJ...936..190W}, $^c$\citet{2019ApJ...886...82R, 2021A&A...649A..75M, 2021PhRvD.104f3043M},     $^d$\citet{2001A&A...368..377B, 2021PhRvD.104f3043M, 2022EPJC...82..778R}, $^e$\citet{2019ApJ...886...82R, 2020EPJC...80..300R, 2021A&A...649A..75M, 2022ApJ...929...56R}, $^f$\citet{2018ApJ...852...53R}, $^g$\citet{2018ApJ...869..101R, 2019ApJ...874...39W, 2020ApJ...893..148R}, $^h$ \citet{2017AdAst2017E...5C,2023ApJ...955...93A}. UPE stands for ultrarelativistic prompt emission, SXFs for soft X-ray flares, HXFs for hard X-ray flares, CED for classical electrodynamics, QED for quantum electrodynamics, SN for supernova, and HN for hypernova. Lower: Examples of the GRB Episodes identified in some BdHNe~I, II, and III. We refer to \citet{2023ApJ...955...93A} and references therein for further quantitative details on each Episode of these sources.}\label{tab:episodes}
    \small\addtolength{\tabcolsep}{-3pt}
    \scriptsize{
   \begin{tabular}{l|c|c|c|c|c|c|ccc|ccc|c}
     \hline
     \multirow{4}{*}{\parbox{2.0cm}{Physical\\phenomenon}} & \multirow{4}{*}{\parbox{0.7cm}{\centering BdHN type}} & \multicolumn{12}{c}{GRB Episodes}\\
     \cline{3-14}
      & & I & \multicolumn{2}{c}{II} \vline & \multicolumn{1}{c}{III}\vline  & IV & \multicolumn{3}{c}{V}\vline  & \multicolumn{3}{c}{VI} \vline& \multicolumn{1}{c}{VII}\\
     &&(SN-rise) &\multicolumn{2}{c}{ ($\nu$NS-rise and NS-rise)} \vline &\parbox{1.5cm}{\centering (BH-rise overcritical)}  & \parbox{1.7cm}{\centering (BH-rise undercritical)} & \multicolumn{3}{c}{(BH echoes)}\vline    & \multicolumn{3}{c}{(Afterglows)} \vline& (SN Ic \& HN)\\
     \cline{3-14}
     & & SN-rise & $\nu$NS-rise  & NS-rise & UPE  & \parbox{1.3cm}{\centering Jetted emission} & Cavity & HXF & SXF & \,\,X\, & Opt. & Rad. & Opt. SN \& HN\\
     & &(X-$\gamma$)&(X-$\gamma$)  & (X-$\gamma$) & (X-$\gamma$) & (GeV) &(X-$\gamma$) &(X-$\gamma$) &(X)  & & & &\\
     \hline
    CO core-collapse$^a$ & I,II,III  & $\bigotimes$&  &  & & & & & & & & &  \\
    \cline{1-2}
    $\nu$NS accretion$^b$ & I,II,III & & $\bigotimes$ & & & & & & & & & & \\
    \cline{1-2}
    NS accretion$^b$  & I,II & & &$\bigotimes$ &  & & & & & & & &\\
    \cline{1-2}
    \parbox{2.3cm}{BH QED$^d$: $e^+e^-$ accel. and transp. (low baryon load)} & I & & & &$\bigotimes$& & & & & & & &\\
     \cline{1-2}
     \parbox{2.3cm}{BH CED$^e$: $e^-$ accel. and radiation} & I & & & & & $\bigotimes$ & & & & & & &\\
    \cline{1-2}
    \parbox{2.3cm}{BH QED$^f$: $e^+e^-$ accel. and transp. (high baryon load)} & I & & & & & & $\bigotimes$& $\bigotimes$ & $\bigotimes$ & & & &\\
     \cline{1-2}
    \parbox{2.3cm}{$\nu$NS synchr. and
    pulsar emission$^g$}  & I,II,III & & & & & & & & &$\bigotimes$ &$\bigotimes$ &$\bigotimes$ &\\
     \cline{1-2}
    \parbox{2.3cm}{Nickel decay and ejecta kinetic energy$^h$} & I,II,II & & & & & & & & & & & &$\bigotimes$\\
    \hline
    \hline
    Source&\parbox{0.7cm}{\centering BdHN type}&\multicolumn{12}{c}{GRB Episodes}\\
    \hline
    GRB 180720B & I & & $\bigotimes$ & $\bigotimes$ & $\bigotimes$ & $\bigotimes$ & $\bigotimes$ & $\bigotimes$ & $\bigotimes$ & $\bigotimes$ & $\bigotimes$ & $\bigotimes$ & $\bigotimes$\\
    GRB 190114C & I & & $\bigotimes$ &  & $\bigotimes$ & $\bigotimes$ & $\bigotimes$ & $\bigotimes$ & $\bigotimes$ & $\bigotimes$ & $\bigotimes$ & $\bigotimes$ & $\bigotimes$\\
    GRB 190829A & II & & $\bigotimes$ & $\bigotimes$ &  &  &  & &  & $\bigotimes$ & $\bigotimes$ & $\bigotimes$ & $\bigotimes$\\
    GRB 171205A & III & & $\bigotimes$ &  &  &  &  & &  & $\bigotimes$ & $\bigotimes$ & $\bigotimes$ & $\bigotimes$\\
    \hline
    \end{tabular}
    }
\end{table*}

\bibliography{sample}{}

\end{CJK*}

\end{document}